\documentclass[aps,preprintnumbers,amsmath,amssymb,nofootinbib, 11pt]{revtex4-1}

\usepackage{setspace}
\pdfoutput=1
\usepackage{graphicx}
\usepackage[T1]{fontenc}
\usepackage{ascii}
\usepackage{cancel}
\usepackage{amssymb}
\usepackage{amsmath,dsfont,textcomp}
\usepackage{bm}
\usepackage{color}
\usepackage{euscript,amsmath}

\DeclareFontFamily{U}{rsfs}{\skewchar\font"7F}
\DeclareFontShape{U}{rsfs}{m}{n}{%
    <-6> rsfs5
    <6-8> rsfs7
    <8-> rsfs10
    }{}
\DeclareMathAlphabet{\mathscr}{U}{rsfs}{m}{n}

\newcommand{\GeV}{~\text{GeV}}
\newcommand{\TeV}{~\text{TeV}}
\newcommand{\BH}{\text{BH}}

\newcommand{\md}{{M_D}}

\newcommand{\mt}{{\tilde{M}}}

\def\lsim{\mathrel{\rlap{\lower4pt\hbox{\hskip1pt$\sim$}}
    \raise1pt\hbox{$<$}}}         
\def\gsim{\mathrel{\rlap{\lower4pt\hbox{\hskip1pt$\sim$}}
    \raise1pt\hbox{$>$}}}         

\begin{document}

\title{ \vspace{1cm} Black holes and the LHC: A review \footnote{Invited review article submitted to {\it Progress of  Particle  and Nuclear Physics.}}}
\author{Seong Chan Park \footnote{s.park@skku.edu}
\\
Department of Physics, Sungkyunkwan University\\
Suwon  440-746 Korea }


\begin{abstract} 
In low-scale gravity models, a  particle collider with  trans-Planckian collision energies can be an ideal place for producing black holes because a large amount of energy can be concentrated at the collision point, which can ultimately lead to black hole formation. In this article, the theoretical foundation for microscopic higher dimensional  black holes is reviewed and the possible production and detection  at the LHC  is described and critically examined.
\end{abstract}%

\maketitle  
\eject

\tableofcontents

\newpage

\section{Introduction} 

A long standing problem in high energy physics is understanding the quantum nature of gravity.  The problem is often defined in terms of the Planck scale, $M_{\rm P}$, which characterizes the strength of the gravitational interaction and sets the cut-off scale of the  effective quantum field theoretic formulations for the strong, weak nuclear and electromagnetic interactions. Over the past several decades, the Planck scale has been regarded as a fundamental scale of physics,  but the highness of the scale has been also  known to  lead to the notorious hierarchy problem. The big hierarchy between the Planck scale ($\sim 10^{19}\GeV$) and the electroweak scale ($\sim 10^3 \GeV$) can be translated into the extreme ratio between the gravitational  constant, $G_N=1/M_{\rm P}^2$, and the Fermi  coupling constant, $G_F$, which represents the relative weakness of the gravitational interaction compared to  the weak nuclear interaction, as follows:
\begin{eqnarray}\label{eq:hierarchy}
\frac{G_N}{G_F}=\frac{6.709 \times 10^{-39} \GeV^{-2}}{1.166 \times 10^{-5} \GeV^{-2}} \sim 10^{-34}.
\end{eqnarray}
The presence of a high scale is problematic particularly in the scalar sector since there is a large quantum quadratic correction,  $\delta m_{\rm 1-loop}^2 \sim \Lambda^2$.  The Higgs sector of the standard model thus suffers from a fine-tuning problem at an extremely high level, %
\begin{eqnarray}
\Delta_{\rm tuning} \equiv \frac{m_{\rm tree}^2 + \delta m_{\rm 1-loop}^2}{\Lambda^2} \sim \frac{(100\GeV)^2}{(10^{19}\GeV)^2}\sim 10^{-34},  
\end{eqnarray}
when $\Lambda \sim M_{\rm P}$. It is evident  that the level of fine tuning, $\Delta_{\rm tuning}$,   is essentially given by the ratio in Eq. \eqref{eq:hierarchy}. The technical aspect of the fine-tuning problem is known to  be  resolved in the presence of additional symmetries such as supersymmetry and conformal symmetry provided that their breaking scales are close to the physical Higgs mass, which still demands additional (dynamical) explanations.  

In the past decade or so a new field has emerged in high energy physics and has expanded to other fields including quantum and classical gravity, cosmology, astrophysics and numerical relativity based on the realization that extra dimensions can lead to ``low scale gravity'' and ``brane world'' models. A higher dimensional brane world  model with large, flat, extra dimensions (Arkani-Hamed, Dimopoulos and Dvali (ADD, 1998) \cite{ADD, ADD2},  Antoniadis+ADD (AADD, 1998) \cite{AADD}) and
a five-dimensional model with a part of highly warped ${\rm AdS}_5$ space (RS (1999) \cite{RS}) have catalyzed an  entirely new approach to the hierarchy problem. 
In these new higher dimensional models, a brane, like an ordinary domain wall, soliton or D-brane in string theory, is introduced to confine a subset or entire fields of the standard model. Gravitons, on the other hand, can propagate through the higher dimensional  bulk and the effective strength of the gravitational interaction for a brane-localized field is diluted by the large volume of extra dimensions or red-shifted by the `warp factor' along extra dimension(s).  As a result, gravity becomes weak as  has been observed in ordinary experiments. The true strength of gravity can only be measured at small distance scales below the compactification radius or the size of extra dimensions, but can cause interesting phenomenological consequences. On the true gravity scale, $M_D$, can be as low as the electroweak scale then the hierarchy problem does not exist. \footnote{To be fair, one should provide the mechanism to stabilize the `radion' or geometry to completely solve the hierarchy problem. See Goldberger-Wise (1999) \cite{Goldberger:1999}  for this mechanism. There is now a wide class of string scenarios, including the ``flux compactifications" (2001) of Giddings, Kachru and Polchinski (2001)\cite{Giddings:2001yu} and of Kachru, Kallosh, Linde, and Trivedi (2003) \cite{KKLT} and other general warped compactifications, which make the distinction between large volume and large warping less clear-cut. One can have either, or a combination of both, in more complete scenarios.}

The most striking prediction in low-scale gravity models is the production and decay of higher dimensional rotating black holes at the CERN Large Hadron Collider (LHC).  Indeed, back in 1974 Penrose first showed that a black hole can be formed in classical high energy head-on collisions \cite{Penrose:1974} and in 1987 't Hooft \cite{'tHooft} made preliminary arguments for graviton dominance and black hole formation in the trans-Planckian domain, which has been clarified by many subsequent studies.  The production of a black hole during particle collision can be easily understood when we accept  the hoop conjecture suggested by Thorne in 1972 \cite{Thorne:1972}.  The hoop conjecture essentially states that a black hole forms if and only if a large amount of energy is packed in a small region that can be surrounded by a hoop with the Schwarzschild radius of the energy. A high energy collider can provide an idealistic environment for this phenomenon as a large amount of energy can be concentrated at the moment of particle collision. In the models with large or highly warped extra dimensions, gravity becomes strong at a low scale in a TeV range so that one can conclude that a collider with a multi-TeV collision energy, such as the LHC,  can produce black holes with ${\rm TeV}^{-1}$ sized event horizons.  Some rough estimation using the geometric cross-section for black hole production, $\sigma \sim \pi r_s(\sqrt{s})^2$,  were made by Banks and Fischler (1999) \cite{Banks:1999}, Giddings and Thomas (2001) \cite{Giddings:2001},  and Dimopoulos and Landsberg (2001) \cite{Dimopoulos:2001} have  drawn considerable attention  not only from physics communities focused on  high energy, string theory, general relativity, numerical relativity, and cosmology, but also from the general public including even  opponents of the LHC experiment (see \cite{safety1} and \cite{safety2} for scientific responses to the opponents).  We also note that Argyres et al. (1998) \cite{Argyres:1998} considered the TeV scale primordial black holes as dark matter candidates or as seeds for early galaxy formation.

After  formation, a black hole radiates energy  through Hawking radiation \cite{Bekenstein:1973, Hawking:1974, Hawking:1976} mainly to the standard model particles on brane \cite{Emparan}.   The decay signatures  depend on greybody factors, ${}_s\Gamma_{\ell m}$, and a thermal factor with angular velocity at the horizon $\Omega$, 
\begin{eqnarray}
\frac{dE_{s,\ell,m}}{dt d\omega} =\frac{1}{2\pi}\frac{{}_s\Gamma_{\ell m}}{e^{(\omega-m \Omega)/T}-(-1)^{2s}},
\end{eqnarray} 
where $\omega, s,\ell,m$ denote the energy, spin and angular momentum quantum numbers of the radiated particle, respectively. In the semi-classical domain,  greybody factors have been obtained for an arbitrary spin ($s=0,1/2,1$) in \cite{IOP1,IOP2-1, IOP2-2, IOP3}, \cite{Harris:2005, Duffy:2005, Casals:2005,Casals:2006} (also see \cite{Casals:2008s, Sampaio:2009tp, Sampaio:2009ra, Kanti:2010mk, Kobayashi:2007zi, Rogatko:2009jp} for various extensions).  The full calculation for gravitational radiation ($s=2$) is still not available,  but there have been some recent developments \cite{graviton1, graviton2, graviton3}.
 Semi-realistic Monte-Carlo (MC) event generators for the black hole events  at the LHC have been developed. The most developed MC generators are BlackMax \cite{Blackmax} and CHARYBDIS \cite{CHARYBDIS, CHARYBDIS2} (also see CATFISH \cite{catfish}). BlackMax v.2 and CHARYBDIS v.2 are used for LHC black hole searches by the CMS and ATLAS collaborations.

In this article, we review the theoretical and experimental foundations of this rapidly developing field. We begin in Section \ref{sec:HighD} with an introduction to physics in higher dimensions, in which we explain low-scale gravity models with large and warped extra dimensions. We also give a short introduction to  compact hyperbolic extra dimensions where  light Kaluza-Klein states are not necessarily expected, but still the gravity scale can be low ($\sim {\rm TeV}$).   Section \ref{sec:blackobjects}
 describes the black hole solutions in higher dimensions where the black hole uniqueness theorem is not generally valid. Emphasis is placed on the Myers-Perry black hole, Black rings and multiple black hole solutions and their stability issues.  Section \ref{sec:energy domain} is devoted to describing physics in the trans-Planckian energy domain. We explain how the known physics of combining quantum field theory and classical general relativity provides good descriptions in the trans-Planckian domain and review the black hole formation process by superimposing the two Aichelburg-Sexl solutions to estimate the cross section for making black holes in particle collisions. An extended phase diagram for the relation between the probing distance and the available energy is also explained at the end.  In Section \ref{sec:LHC} we focus on the LHC experiment to identify black hole signatures.  The expected signatures  are obtained considering the precise greybody factors. Issues in the balding and Planck phases of the produced black hole are discussed. After reviewing the recent LHC result with $\sqrt{s}=7$ TeV collision energy, we critically re-examine the interpretations.  In Section \ref{sec:conclusion} we conclude with a discussion of future higher energy runs with $\sqrt{s}=14 (100) \TeV$ and open problems. We present the relevant physical scales in Appendix \ref{Appendix:1} and briefly sketch the numerical method to solve the generalized Teukolsky equation to calculate the greybody factors in Appendix \ref{Appendix:2}.

Some aspects of this topic have been described in other reviews by Landsberg \cite{review:Landsberg}, Kanti \cite{review:Kanti}, Giddings \cite{review:Giddings} and Park \cite{Park:2008} and also in lectures by Kanti \cite{lecture:Kanti}.



\section{Physics in higher dimensions \label{sec:HighD}}

In this section we first provide a generalization of spacetime taking $n$-extra dimensions into account.  Indeed, no particularly compelling theoretical reason exists for three spatial dimensions.  One may note that only in $(3+1)$ dimensions, the Kepler problem with the potential of the form $V = -\alpha/r$ allows closed orbits \cite{Bertrand:1873} and thus one may find an anthropic reason for the three
existing {\it large} spatial dimensions. However, it still allows the existence of the extra {\it microscopic} dimensions, which are not responsible for the planetary motions. More importantly, consistency conditions for the known extension of Einstein gravity and the quantum field theoretical embedding of it, namely, string and supergravity theories, require a certain number of extra dimensions. Furthermore, it has been revealed that extra dimensions can provide  totally new insights to reformulate the notorious old problems in particle physics including the big hierarchy problem, which will be discussed in this section.

\subsection{Einstein-Hilbert action in higher dimensions}

In $(3+1)$ dimensions, pure gravity is described by Einstein-Hilbert action:
\begin{eqnarray}
S_{\text{EH}}=- \int d^4 x \sqrt{g} ~\frac{1}{16\pi G}R,
\label{eq:EH4}
\end{eqnarray}
where $g=|\text{Det}(g_{\mu\nu})|$. Setting the action dimensionless ($\hbar=1, c=1$),  the Newton's constant $G$ has a dimension of $\left[\text{Length}\right]^2=\left[\text{Mass}\right]^{-2}$ and  thus defines a mass scale, $1/\sqrt{G}$. 
This scale is nothing but the Planck scale $M_{\rm P}=1/\sqrt{G}=1.22089(6) \times 10^{19}$ GeV \cite{PDG}.

In the presence of the extra $n$-dimensions, the Einstein-Hilbert action can be generalized to a form
\begin{eqnarray}
S_{\text{EH}}=S_4 \to S_D = -\int d^D x~ \sqrt{g_D} \frac{1}{16\pi G_D} R_D,
\label{eq:EHD}
\end{eqnarray}
where $G_D$ has a dimension of $\left[\text{Length}\right]^{D-2}=\left[\text{Mass}\right]^{-D+2}$ as $\left[R_D\right]=\left[\text{Length}\right]^{-2}$. A higher dimensional generalization of Einstein equations could be obtained by varying the action in Eq. \eqref{eq:EHD}. One may define a generalized Planck scale in $D$-dimensions (see Appendix \ref{Appendix:1} for Planck units in $D$ dimensions):
 \begin{eqnarray}
\md \equiv \left(\frac{{\cal N}_n}{8\pi G_D}\right)^{\tfrac{1}{2+n}},
 \end{eqnarray}
where ${\cal N}_n$ is a numerical factor for which different definitions have been employed in the literature. Some representative definitions that are widely utilized include the Particle Data Group convention (PDG) \cite{PDG}, which is the same convention in Ida-Oda-Park (IOP) \cite{IOP1}, the Randall-Sundrum convention (RS) \cite{RS}, Giddings-Thomas convention (GT)\cite{Giddings:2001} and Dimopoulos-Landsberg convention (DL) \cite{Dimopoulos:2001}:  
\begin{eqnarray}
{\cal N}_n &=&\left\{\begin{matrix}
(2\pi)^n & \text{PDG , IOP }\\ 
1 & \text{RS}\\ 
2\cdot (2\pi)^n & \text{GT}\\ 
8\pi & \text{DL}.
\end{matrix}\right.
\label{eq:GD}
\end{eqnarray}
In this review, we follow the PDG convention. Now, starting from the higher dimensional action, $S_D$,  the four-dimensional action $S_4$ can be derived as an effective action after integrating out the extra coordinates.  In this respect,  $\md$ should be regarded as a fundamental scale of gravity, from which the Planck scale in four dimensions is derived after compactification.

Let us consider a pretty general model of an extra dimension with a ``warp factor'' $W(y^i)$, which  is a monotonic positive function of the extra coordinate $y^i$.  If we require the 4D Poincar\'e invariance at each point of $y^i$, the full metric of $4+n$-dimensional spacetime can be written in general as
\begin{eqnarray}
ds^2 = W^2(y) \eta_{\mu\nu} dx^\mu dx^\nu + g_{mn}(y) dy^m dy^n,
\label{eq:warp}
\end{eqnarray}
where we assume $y^i\in [0, \ell]$. We do not consider a negative warp factor value to avoid a naked singularity in extra dimensions. With the assumed monotonic property of the warp factor, we can link the change in the position along the extra dimension and the renormalization group ``running" with respect to the energy scale.  Thus one can impose the geometric meaning on the UV-scale and IR-scale in extra dimensions.

\subsection{Flat, factorizable geometry}

If the warp factor is trivial, $W(y)=1$, and the extra dimension is flat, $g_{mn}=-\delta_{mn}$, the resultant geometry is simple, thus we can easily compute the $4D$ effective action as follows:%
\begin{eqnarray}
S_D &\xrightarrow[W(y)=1]{g_{mn}=-\delta_{mn}}& -\int d^n y \sqrt{g_y} \frac{1}{16\pi G_D} \times \int d^4 x \sqrt{g} R + \cdots,  \nonumber \\
&=&- \frac{V_n}{16\pi G_D} \times \int d^4 x \sqrt{g} R +\cdots,
\end{eqnarray}
where we can find the relationship between the Newton's constant in $4D$ and in general dimensions:%
\begin{eqnarray}
G_4 = \frac{G_{4+n}}{V_n}.
\end{eqnarray}
Here, the volume of the extra dimension is denoted by $V_n = \int d^n y \sqrt{g_y}$. If the volume of the flat extra dimension is {\it large}, $V_n \to \infty$, we get a tiny Newton's constant $G_4\to 0$, thus the huge Planck mass $M_4 \to \infty$. The large extra dimensions  provide a geometrical explanation for why we have weak gravity or a large Planck scale as was suggested by (A)ADD \cite{ADD, AADD}.

A particularly interesting possibility is that $M_D$ is as low as the electroweak scale of the order of $1$ TeV ($M_D\sim 1{\rm TeV}$)  such that there is no hierarchy between the fundamental scale and the electroweak scale. The solution to the hierarchy problem works if%
\begin{eqnarray}
\frac{M_4^2}{\md^2} =\frac{1}{(2\pi)^n} \left(\md^n V_n\right)\sim 10^{30} \\
\text{or}\,\,\,\,\,  V_n^{1/n} \sim (2\pi) \times 10^{30/n} \times (1/\md) \sim 10^{\tfrac{30}{n}-3} \text{fm} \nonumber
\label{eq:KK in ADD}
\end{eqnarray}
where  we took $M_4 = 1/\sqrt{8\pi G}\approx 2.4 \times 10^{15}$ TeV and $ 1/\md \sim 1/\TeV \sim 2\times 10^{-4}~ \text{fm}$. The actual numbers for various dimensions are
\begin{eqnarray}
\ell_c = V_n^{1/n} \sim 
\left\{\begin{matrix}
10^{12}  ~\text{m}& n=1, \\ 
1 \text{mm} & n=2,\\ 
10 ~\text{nm} & n=3,\\ 
4\times 10^4 ~\text{fm} & n=4,\\ 
10^3 ~\text{fm} & n=5, \\ 
100 ~\text{fm} & n=6,\\ 
20 ~\text{fm} & n=7.
\end{matrix}\right.
\end{eqnarray}
Obviously the first case, $n=1$, is ruled out by astronomical measurements. The second case, $n=2$, is also ruled out by the short-distance Torsion pendulum experiments carried out by Kapner et al. (2006) \cite{Kapner:2006} and Adelberger et al. (2006) \cite{Adelberger:2006}.  Due to the mass gap, ($\Delta m \sim 1/V^{1/n}$) is small with a large volume, and Kaluza-Klein (KK) gravitons can contribute to astrophysical processes such that models with a larger number of extra dimensions ($n\geq 3$) can also be strongly constrained. The requirement that KK gravitons do not contribute to the supernova cooling process by carrying the energy away gives the bound $\md >$ a few TeV for $n=3$ (Hanhart et al. 2000, 2001)  \cite{Hanhart:2000, Hanhart:2001}. An even stronger bound, $\md>4$ TeV  was found by Hannestad-Raffelt (2001) \cite{Hannestad:2001} from measurements of diffuse $\gamma$ ray detection by the EGRET satellite for $n=3$  and also for more involved processes with supernovas, neutron stars and pulsars (Hannestad-Raffelt (2002)) \cite{Hannestad:2002}.  All these astrophysical bounds are for  light KK gravitons $\Delta m < 100$ MeV,  thus a nontrivial extra dimension model with a larger mass gap is not constrained by them.   Taking the possible large astronomical uncertainties into account, $n\geq 3$ cases are still not excluded.


\subsection{Non-factorizable geometry}

When $W(y)$ is a non-trivial function of $y$ (again we still require a monotonic property), the geometry is warped and the effective theory has a scaling behavior, $\Lambda(y) = W(y) \Lambda$. 
As in the flat case, we can read out the effective action by integrating the extra dimensions out as follows:
\begin{eqnarray}
S_D &\xrightarrow[W(y)]{g_{mn}=-\delta_{mn}}& -\int d^n y  W^4(y) \frac{1}{16\pi G_D} \times \int d^4 x \sqrt{\bar{g}} W^{-2}\bar{g}^{\mu\nu}R_{\mu\nu} + \cdots,  \nonumber \\
&=&- \frac{\tilde{V}_n}{16\pi G_D} \times \int d^4 x \sqrt{\bar{g}} \bar{R} +\cdots,
\end{eqnarray}
where barred quantities are canonically normalized. We  can derive the conventional Newton constant
out of the higher dimensional Gravitational constant as
\begin{eqnarray}
G_4 = \frac{G_{4+n}}{\tilde{V}_n},
\end{eqnarray}
where ``warped volume'' is defined as
 \begin{eqnarray}
 \tilde{V}_n = \int d^n y \sqrt{g_y} W^2(y)
 \end{eqnarray}
 with the nontrivial warp factor $W(y)$ in the metric along the extra dimensions.

 For concreteness,  we consider the $n=1$ case in detail here
\begin{eqnarray}
M_4^2 = M_5^3 \frac{\tilde{V}_1}{2\pi},
\end{eqnarray}
which yields a natural relationships among dimensionful parameters, $M_4 \sim M_5\sim 1/\tilde{V}_1$. For instance, an AdS space with  $W(y) = e^{-k y}$ gives $\tilde{V}_1 = \int_0^\ell  dy e^{-2 ky} = \frac{1}{2k}\left(1-e^{-2k\ell}\right)\sim \frac{1}{2k}$ where $k$ is the AdS curvature. Thus,  $k\sim M_5\sim M_4$.

Notably the standard model (or at least the Higgs field) is confined to the ``IR-brane" located at $y=\ell$.  (Conversely, the ``UV-brane" is at $y=0$.) The Lagrangian for the Higgs field is read as follows:
\begin{eqnarray}
S_{\text{Higgs}}=\int d^4 x  \int_0^\ell dy \delta(y-\ell) W^4(y) {\cal L}\left(W^2(y)\eta_{\mu\nu}, H\right)\\
=\int d^4 x W^4(\ell) {\cal L}\left(W^2(\ell)\eta_{\mu\nu}, H\right).
\end{eqnarray}
By rescaling the kinetic term in ${\cal L}$ and taking the warp factor into account, the physical Higgs field with the canonically normalized kinetic term is defined as $H_{\text Phys}=W(\ell) H$, and the physical mass parameter on the IR (and UV) brane is given as 
\begin{eqnarray}
&&M_{\text{Phys}}(\ell)=W(y_{\rm IR}=\ell)M,\\
&&\frac{M_{\rm Phys}(y_{\rm UV})}{M_{\rm Phys}(y_{IR})}=\frac{W(y_{\rm UV})}{W(y_{\rm IR})}\gg 1.
\end{eqnarray}
In AdS space, $M_{\text{Phys}} =e^{-k\ell} M \ll M$ such that the actual physical mass scale on the IR-brane is exponentially small compared to the one on the UV-brane. This phenomenon takes place due to the warp factor, which changes the ``ruler'' at each position $y$. Having exponential dependence, we can easily generate a big hierarchy between the UV-scale and the IR-scale as Randall and Sundrum (1999) \cite{RS} showed.  A slice of AdS space arises as a solution to the five-dimensional Einstein equations with a negative cosmological constant (AdS curvature) in the bulk and well-balanced tensions for UV (IR)-branes. The setup provides an explanation as to why the IR-scale (or electroweak scale) is much lower than the UV scale (or Planck scale). In this setup, the fields on the IR-brane can ``feel'' the strong gravity  on the IR scale $M_{\text{IR}}\sim e^{-k\ell} M_{\text{UV}} \sim e^{-k\ell} M_5$, and the Planck-TeV hierarchy is given by the warp factor:
\begin{eqnarray}
\frac{M_{\rm IR}}{M_{\rm UV}}\sim 10^{-15} = e^{-k\ell}.
\end{eqnarray}

The most important lesson here is that gravity may become strong at the IR scale ($\sim \TeV$)  in the presence of warped extra dimensions. If the extra dimension is large or highly warped, the strong gravity scale can be as low as the electroweak scale such that there is no hierarchy between the gravity scale and the electroweak scale. This is a unique feature of a  higher dimensional setup that cannot be found in four-dimensional effective theories. This understanding opens up an entirely new area of extremely interesting phenomenological consequences including  black hole production at the LHC.

\subsection{Compact hyperbolic space \label{Sec:CHS}}

An intriguing possibility is that we live in higher dimensional spacetime, which contains compact hyperbolic extra dimensions. The compact hyperbolic spaces (CHS) have been relatively less intensively studied by physicists  so far but
its interesting mathematical properties have been explored since original work by
distinguished mathematicians including  Selberg(1965) \cite{Selberg}, Mostow(1968) \cite{Mostow} and Thurston (1978)\cite{Thurston}. In particular, there are two properties listed below that deserve further attention. 

\begin{itemize}
\item {Rigidity}: Proved by Mostow (1968).
\item {Large mass gap}: Conjectured by Selberg (1965).
\end{itemize}

A CHS can be obtained by taking a quotient space of a $d-$hyperbolic space (${\cal H}^d$) with its fundamental group ($\Gamma$): ${\cal H}^d/\Gamma$. If we do not consider a  cusped hyperbolic three-manifold or incomplete metric, the {\it Mostow's rigidity theorem} states that the geometry of a CHS of dimension greater than two ($d>2$) is determined by its fundamental group. This means that once we fixed the local curvature of CHS, $\ell_c$,  and the volume, no more moduli remains to be fixed.  This is largely different from other extra dimensions which often suffer from shape moduli stabilization issues. Of course, there remains the issue of super-selection of a particular fundamental group among many others. 

The large mass gap of CHS is based on the conjecture by Selberg (1965), who stated that  there is a lower bound of the first eigenmode of the Laplace-Beltrami operator on CHS determined by the following inequality: 
\begin{eqnarray}
k_1 \ell_c \geq \frac{1}{2}.
\end{eqnarray}
Indeed Luo and Rudnick (1995) \cite{Luo} proved the conjecture up to $k_1 \ell _c\geq \sqrt{171/784} \approx 0.22$ using number theory approaches,  and 
Cornish and Turok (1998) \cite{Cornish} and Cornish and Spergel (1999) \cite{Cornish:1999} also  studied  higher excitation modes on CHS. 

The presence of the lower bound  opens up an entirely new perspective of extra dimensional physics. Usually, as in torus compactification, a large volume implies a small mass gap of the order of $1/{\text{Vol}_n}^{1/n}$. Since the large volume is responsible for the small fundamental gravity scale (see Eq. \eqref{eq:KK in ADD}), a small mass gap or a light KK excitation of a bulk field seems unavoidable. However, due to the existence of the Selberg's bound, CHS does not have to accompany a light mode. This is a big advantage in terms of model building. If there is no light Kaluza-Klein state, the  effect of extra dimensions is hidden and the model  does not affect low energy experiments. In this case, the extra dimension is evident only through the strong gravity effect such as black hole production at particle colliders.  This possibility was recently studied  by Orlando and Park (2010) \cite{Orlando:2010} (also see \cite{Kim:2010} for an inflationary model based on CHS).

\subsection{Summary and discussion}

In pure general relativity, the only input parameter is the number of spacetime dimensions. In string theory, a certain number of extra dimensions are required to be introduced to formulate the theory in a consistent way. Recently we have observed in many theoretical studies that the presence of extra dimensions provides new insights into many problems in particle physics too. 

On the other hand, we do not have any concrete experimental `data' for extra dimensions thus we  can  only theoretically model the topological and geometrical structure of `hidden' dimensions. Recently suggested `large' and `highly warped' and also `hyperbolic' extra dimensions have opened  a new window to addressing the problem of weak gravity or hierarchy in energy scales. A particularly interesting possibility is that the fundamental scale of gravity (which is supposed to be the largest scale of physics) may not be extremely large as was traditionally suggested by Planck energy, but can be as low as a TeV at least for a Higgs sector. If the low scale of gravity is really built in, we may learn about the presence of extra dimensions by experimentally probing the strong gravitational phenomena. The most dramatic phenomenon, which may not be  that far from us, is microscopic black hole production and their decay at the LHC and sky \cite{Anchordoqui:2003astro}. Most probably those microscopic black holes are higher dimensional as long as their `size' is much smaller than the compactification radius or local curvature radius of the background extra dimension:
\begin{eqnarray}
r_{\rm BH} \ll r_c.
\end{eqnarray}

However, one should note that the details of `black hole signatures'  depend on the detailed structure of extra dimensions and also the precise distribution of matter fields in higher dimensions which we can only make models. Also we have to assume that our theoretical understanding of the quantum nature of gravity is limited. The best thing we can try now is making concrete and sound `predictions' of a given set of models with theoretical confidence. A black hole can be well described by a semi-classical picture as long as its `size' is big compared to the Planck length:
\begin{eqnarray}
\ell_D \ll r_{\rm BH}.
\end{eqnarray}
The next section discusses for such semi-classical and higher dimensional black holes.

\newpage

\section{Black objects in higher dimensions \label{sec:blackobjects}}

In this section, we  review the black objects in higher dimensions. Indeed in higher dimensions, a generalization of the black hole uniqueness theorem is not valid \cite{Gibbons:2002a, Gibbons:2002b, Gibbons:2002c, Rogatko:2002, Rogatko:2003, Morisawa:2004}, and a number of black objects can exist that share the same mass and angular momentum. 
For instance, the higher dimensional generalization of the Kerr black hole, or the Myers-Perry(MP) black hole, can share the same mass and angular momentum with black rings (BR), the topology of which is distinguishable from the simple sphere of the MP black hole: $S^2\times S^1$ (BR) vs. $S^3$ (MP bh) in five dimensions ($D=5$). An even further set of solutions seem to exist in higher dimensions ($D>5$), thus the {\it classification} of all stationary, asymptotically flat (or AdS, dS or hyperbolic) black hole solutions to the higher dimensional Einstein equation (in vacuum or with coupled matter) becomes a major problem in the study of higher dimensional gravity. 

In particle collisions, however, the most relevant solution is  the MP black hole as other solutions seem to suffer from various instabilities (there is instability even for the MP black hole, especially at a high rotation limit). In this section, we focus on the MP black hole, black rings and other black hole solutions in higher dimensions.

\subsection{Schwarzschild-Tangherlini solution in higher dimensions}

The generalization of the Schwarzschild solution in higher dimensions is obtained by replacing $1/r$ dependence of the Newtonian potential with  $1/r^{D-3}$ as Tangherlini found in 1963 \cite{Tangherlini:1963}
: 
\begin{eqnarray}
ds_{\rm Sch}^2=\left(1-\frac{\mu}{r^{D-3}}\right) dt^2 -\frac{dr^2}{1-\frac{\mu}{r^{D-3}}}-r^2 d\Omega_{D-2}^2,
\end{eqnarray}
where  a mass parameter $\mu$ is defined as
\begin{eqnarray}
\mu =\frac{16\pi G_D M}{(D-2)\Omega_{D-2}},
\end{eqnarray}
where $\Omega_n = \frac{2\pi^{(n+1)/2}}{\Gamma((n+1)/2)}$  is the area of a unit $n$-sphere.
Under the linearized gravitational perturbations, the $D\geq 4$ Schwarzschild solution is stable as Ishibashi and Kodama showed in 2003 \cite{Ishibashi:2003, Kodama:2003}.

\vspace{.5cm}
\begin{figure}[h]
\begin{center}
        \includegraphics[width=.45\textwidth]{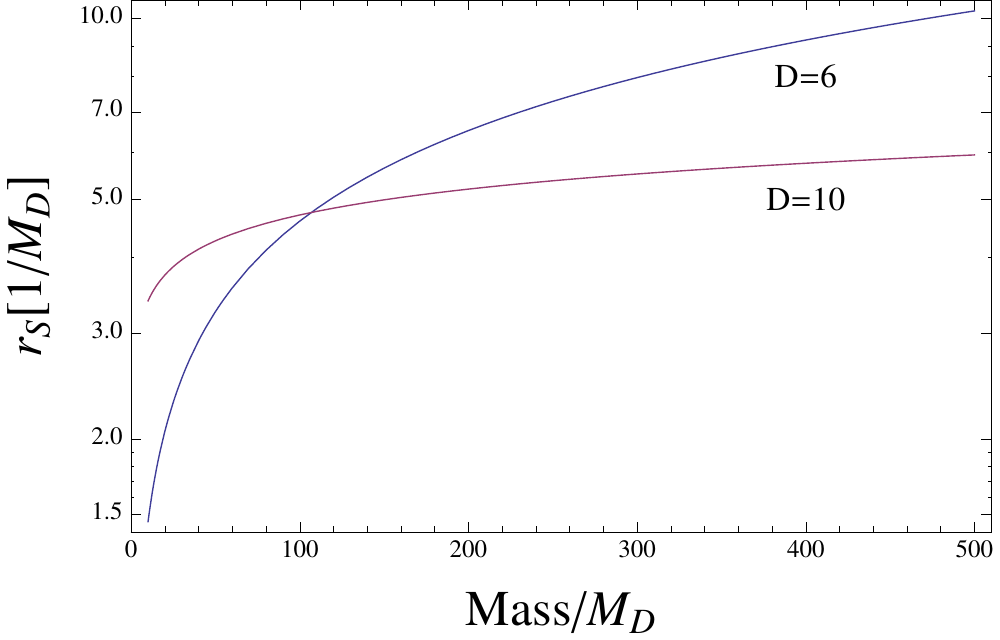}
        \end{center}
  \caption{The size of the Schwarzschild Black hole in $D=5,6, \cdots, 11$ dimensions for a range of masses.
 One should notice that a black hole can be described by a classical GR only when $r_s\gg 1/\md$ such that one can neglect the quantum gravitational corrections.}
\label{fig:radius}
\end{figure}

The event horizon locates at  $r^{D-3}=\mu$:
\begin{eqnarray}
r_s &=& \mu^{\tfrac{1}{D-3}} \\
&=& \left(\frac{16\pi G_D M}{(D-2)\Omega_{D-2}}\right)^{\tfrac{1}{D-3}} \nonumber\\
&=& k_{D-4} \left(\frac{M}{\md}\right)^{\tfrac{1}{D-3}}\frac{1}{\md}, \,~~~~~
k_n=\left(\frac{{\cal N}_n \Gamma(\tfrac{3+n}{2})}{(n+2)\pi^{(n+3)/2}}\right)^{\tfrac{1}{n+1}},
\label{eq:rs}
\end{eqnarray}
where ${\cal N}_n$ is given in Eq. \eqref{eq:GD} for $n(=D-4)$ extra dimensions. Explicitly, using the PDG convention where ${\cal N}_n =(2\pi)^2$, we found
\begin{eqnarray}
k_n =
\left\{\begin{matrix}
1/4\pi\approx 0.08  & n=0, \\ 
(2/3\pi)^{1/2}\approx 0.46 & n=1,\\ 
(3/4)^{1/3}\approx 0.90 &  n=2,\\ 
(2/5^{1/4})\approx 1.33 & n=3,\\ 
(5\pi)^{1/5}\approx 1.73 & n=4, \\ 
2\left(3\pi/7\right)^{1/6} \approx 2.1& n=5,\\ 
(105 \pi^2/2)^{1/7}\approx 2.4  & n=6, \\
2(2\pi)^{1/4}/3^{1/8}\approx 2.76 & n=7.
\end{matrix}\right.
\label{eq:kn}
\end{eqnarray}
Since the Schwarzschild radius increases as $\sim M^{1/(D-3)}$ the size slowly increases with respect to the mass  in higher dimensions (see Fig. \ref{fig:radius}).

The Hawking temperature and the horizon area (or entropy), respectively, are given as
\begin{eqnarray}
T_h &=&  \frac{D-3}{4\pi r_s},\\
S_h &=& -\frac{1}{D-2}\frac{\partial M}{\partial T} = \frac{r_s^{D-2}\Omega_{D-2}}{4G_D}.
\end{eqnarray}
As $M$ increases, $r_s$ and $S_h$ also increase,  but $T_h$ decreases. The smallest black hole is the hottest. 

\begin{figure}[h]
\begin{center}
\includegraphics[width=.45\textwidth]{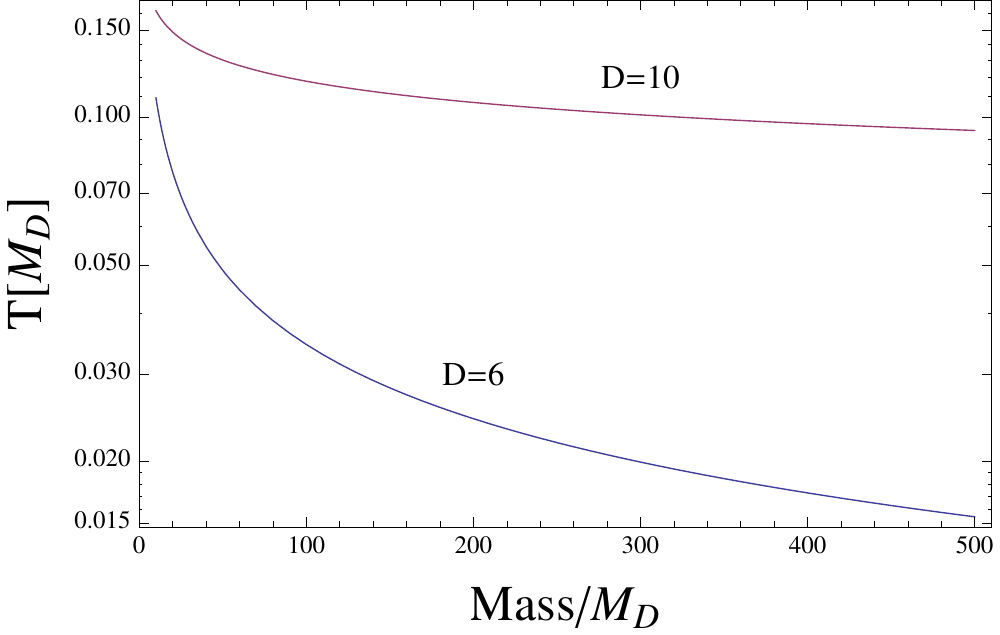}\\
        \includegraphics[width=.45\textwidth]{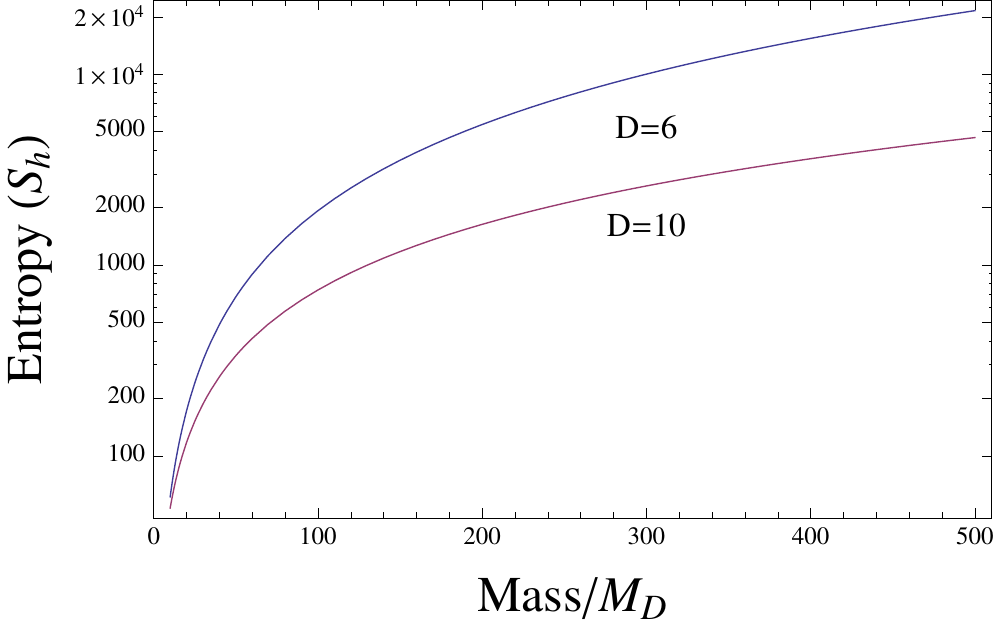}                
        \end{center}
  \caption{The thermodynamic variables (temperature, entropy) of the Schwarzschild black hole in $D=5$ and $D=10$ dimensions, respectively.}
\label{fig:thermo}
\end{figure}

\subsection{Myers-Perry black holes}

In $D$-dimensions the number of independent rotation planes is $N=\left[(D-1)/2\right]$.   In 4D, a unique rotation plane exists, but in 5D and 6D, for instance, there are two independent planes. In particle collisions, however, an angular momentum vector, which is perpendicular to the collision plane of the particles, is given so that a black hole solution with one angular momentum is 
the most relevant among all possible solutions.

The higher dimensional rotating black hole solution was found by Myers and Perry (1986) \cite{Myers-Perry}. With one rotation, which is parametrized by a parameter $a$, the solution takes the form
\begin{eqnarray}
ds_{\rm MP}^2= dt^2 -\frac{\mu}{r^{D-5}\Sigma} (dt - a s_\theta^2 d\phi)^2 -\frac{\Sigma}{\Delta} dr^2
-(r^2+a^2)s_\theta^2 d\phi^2-r^2c_\theta^2 d\Omega_{D-4}^2,
\end{eqnarray}
where we used shortened  notations $s_\theta=\sin\theta$, $c_\theta=\cos\theta$ and
\begin{eqnarray}
\Sigma &=& r^2+a^2 c_\theta^2, \\
\Delta &=&  r^2+a^2 -\mu/{r^{D-5}}.
\end{eqnarray}
The physical  ADM mass ($M$) and angular momentum ($J$), which are obtained by comparing the metric at $r=\infty$ with the metric for a massive perturbation in Minkowski spacetime, are, respectively related to the $\mu$ and $a$ parameters: 
\begin{eqnarray}
M &=& \frac{(D-2)\Omega_{D-2}}{16\pi G_D}\mu,\\
J &=& \frac{2}{d-2} M a,
\label{eq:MJ}
\end{eqnarray}
where $G_D$ is Newton's constant in $D$-dimensions (See Eq. \eqref{eq:GD}). The $a\geq0$ parameter is the angular momentum per unit mass, which is common in the literature. The event horizon for $D>4$  is set at $\Delta(r=r_H)=0$ or
\begin{eqnarray}
r_h=\left(\frac{\mu}{1+a_*^2}\right)^{\tfrac{1}{D-3}} \equiv \frac{r_s}{(1+a_*^2)^{\tfrac{1}{D-3}}} \leq r_s,
\label{eq:rh}
\end{eqnarray}
and the area is given as
\begin{eqnarray}
A_H =r_h^{D-4} (r_h^2+a^2) \Omega_{D-2},
\end{eqnarray}
where a dimensionless rotation parameter $a_* = a/r_H$ is recursively defined. $r_s$ is the Schwarzschild radius for the corresponding mass $M$ and the angular momentum $J=0$.

 A higher dimensional Schwarzschild solution (i.e., spherically symmetric, static black hole solution) is recovered when $a=0$. 
In this {\it non-rotating} limit, 
\begin{eqnarray}
\Sigma &\to& r^2, \\
\Delta &\to& r^2 - \mu/r^{D-5}, \\
J&\to& 0,\\
\mu &\to& r_s^{1/(D-3)}.
\end{eqnarray}

An upper bound on $a$ exists for $D=4$  where
\begin{eqnarray}
&&\Delta (r_h) = r_h^2 -\mu r_h + a^2 = (r_h - \frac{\mu}{2})^2 + a^2-\mu^2/4 =0, \\
&&a^2 = \mu^2/4 -(r_h-\mu/2)^2 \leq \left(\mu/2\right)^2,\\
&&\therefore  a \leq \mu/2.
\end{eqnarray}
Similarly for $D=5$,
\begin{eqnarray}
&&\Delta(r_h)=r_h^2 -\mu +a^2 =0, \\
&&r_h^2 = \mu -a^2 \geq 0, \\
&& \therefore  a \leq \sqrt{\mu},
\end{eqnarray}
or $\frac{J^2}{M^3} \leq \frac{32G}{27\pi}$ for $D=5$.
In contrast, for $D>5$, there is no upper limit on $a$ so that, in principle, a black hole can rotate with an arbitrarily large $a$. These highly rotating black holes are often called {\it ultraspinning}, and are  known to be unstable as shown by Emparan and Myers (2003) \cite{Emparan:2003us}.
Recently, a bar mode instability was reported by Shibata and Yoshino (2010) \cite{Shibata-Yoshino:2010}.

The Hawking temperature and the angular velocity and the horizon area (or entropy), respectively,  are given as
\begin{eqnarray}
T_h &=& \frac{(D-3)+(D-5)a_*^2}{4\pi (1+a_*^2)r_h} \to \frac{D-3}{4\pi r_s},\\
\Omega &=&\frac{a_*}{(1+a_*^2) r_h} \to 0,\\
S_h &=& -\frac{1}{D-2}\frac{\partial M}{\partial T} =\frac{M}{(D-2)T_h}(n+1-\frac{2a_*^2}{1+a_*^2}) \to \frac{r_s^{D-2}\Omega_{D-2}}{4G_D}.
\end{eqnarray}
where the limiting cases correspond to $J\to 0$ or a non-rotating limit.


\subsection{Other higher dimensional black holes}

\subsubsection{Black strings and black p-branes}

Since a direct product of Ricci flat spaces is also Ricci flat, a new class of vacuum solutions can be constructed in $D+p$ dimensions out of a vacuum black hole solutions with event horizons in $D$ dimensions. Explicitly, a black p-brane solution is obtained by a direct product of the Schwarzschild black hole and a flat manifold:
\begin{eqnarray}
ds^2_{Bp-brane}= ds^2_{Sch} \times \sum_i^p \delta_{ij}dy^i dy^j,
\end{eqnarray}
in which the event horizon of the Schwarzschild black hole is extended in $y^i$ directions. Thus the horizon topology is from ${\cal H}$ to ${\cal H}\times {\rm R}^p$. If the $y^i$ directions are compact by periodic condition $y^i \sim y^i +2\pi R_i$, the horizon topology becomes ${\cal H}\times {\rm T}^p$. If  $p=1$, in particular, the new black hole solution is called black string.  Having an extended horizon, the black p-branes are not  asymptotically flat. A `long' black string is known to be unstable as Gregory-Laflamme showed in 1993 \cite{Gregory:1993} and 1994 \cite{Gregory:1994} (also see Harmark, Niarchos and Obers (2007) \cite{Harmark:2007md} and Cardoso and Dias (2004) \cite{Cardoso:2004} for instability of a small Kerr-AdS black hole).

\subsubsection{Black rings}
In addition to the  MP solutions, another black hole solution with horizon topology $S^2 \times S^1$ (or donut) was found by Emparan and Reall in 2002 \cite{Emparan:2001wn}. The asymptotic geometry is flat. The $S^1$ describes a circle that is stabilized by the centrifugal force of the rotating ring so that  a lower limit on the rotating parameter exists different from the MP black hole. The solution is given in a form \cite{Emparan:2004wy}
\begin{eqnarray}
ds_{\rm BR}^2 = \frac{F(y)}{F(x)}(dt - C R \frac{1+y}{F(y)} d\psi)^2 -\frac{R^2}{(x-y)^2}F(x) \left(-\frac{G(y)}{F(y)} d\psi^2 - \frac{dy^2}{G(y)} +\frac{dx^2}{G(x)} +\frac{G(x)}{F(x)} d\phi^2 \right),
\end{eqnarray}
where $F(x)=1+\lambda x$, $G(x)=(1-x^2)(1+\nu x)$ and  $C=(\lambda(\lambda-\nu)(1+\lambda)/(1-\lambda))^{1/2}$. The parameters $\lambda$ and $\nu$ vary in the range $1>\lambda \geq \nu >0$, the coordinates $-1\geq y \geq -\infty$ and $1\geq x \geq -1$. The axis of rotation around the $\psi$ direction is at $y=-1$. The other rotation axis around $\phi$ is $x=1$ and $x=-1$ inside and outside of the ring, respectively. The horizon is at $y=-1/\nu$. 

In five dimensions, the spin can also grow indefinitely, but only if the spinning object is a ring:
\begin{eqnarray}
\frac{J^2}{M^3}>0.8437 \times \left(\frac{32G}{27\pi}\right),
\end{eqnarray}
where $\frac{32G}{27\pi}$ is the upper limit of the MP black hole in $D=5$.
Due to the existence of the lower limit on the angular momentum, we may conclude that black rings may not be produced in particle collisions  as discussed by Ida, Oda and Park (2003 )\cite{IOP1}. However 
the production of black rings is not conclusive in $D>5$.

\subsubsection{Multi black holes}
In $D=4$ there are no  stationary vacuum multi-black hole solutions to the Einstein equations. However, such black holes exist in $D=5$ and also in  $D>5$. These multi black hole configurations could not be obtained by initial data about colliding particles, but they have their own interesting issues (see however, Yoshino-Shiromizu \cite{Yoshino:2007b} for black ring formation by closed string collisions).

The simplest multi black hole solution in higher dimensions is `black saturn', which was constructed by Elvang and Figueras in 2007 \cite{Elvang:2007}.  Black saturn is  composed of a Myers-Perry black hole in the center, which is geometrically surrounded by a black ring. The angular momentum of the ring makes two black holes remain apart and stay in balance.  An interesting property of black saturn is that its total angular momentum measured at infinity can vanish. This can happen when the black ring and the central MP black hole rotate in opposite directions. This is very distinctive from the four-dimensional case for which the Schwarzschild solution is the unique solution in $D=4$ with vanishing angular momentum ($J=0$). Other more complicated configurations are also known such as the black di-ring discussed by Iguchi and Mishima (2007) \cite{Iguchi:2007} and bicycling black rings by Elvang and Rodriguez (2008) \cite{Elvang:2007B}. Multiple-MP black hole configurations have also been  considered, but the existence of conical singularities seems to be unavoidable \cite{Herdeiro:2008}. A good review of exact solutions for higher dimensional black holes was recently presented by Emparan and Reall (2008) \cite{review_Emparan:2008} and  Tomizawa and Ishihara (2011) \cite{Tomizawa:2011}.

\subsection{Summary}
It is important to realize that the black hole uniqueness theorem is generally not valid.  The different black objects in higher dimensions can share their masses and angular momenta even though their geometrical properties are completely different. However, among many new  black hole solutions that  have been constructed in higher dimensions to date,  the Myers-Perry(MP) and Schwartzschild-Tangherlini (ST) solutions seem to provide the best realistic descriptions for rotating and non-rotating black holes at the LHC. Indeed  the MP and ST solutions have classical stability under linearized gravitational perturbations so that they are the most probable candidates for produced black objects in particle collisions.  Keeping this fact  in mind, we summarize the most relevant physical quantities for the ST black hole ($J=0$ limit of MP black hole) in various dimensions for future reference. We introduce a convenient dimensionless parameter $\tilde{M}= M/ \md$ for the black hole mass $M$ and explicitly write $r_s, T_h$ and $S_h$ in various dimensions:

\begin{eqnarray}
r_s \md &=&
\left\{\begin{matrix}
(2/3\pi)^{1/2} (M/\md)^{1/2} \approx 0.46 \mt^{1/2}  & n=1\\ 
(3/4)^{1/3} (M/\md)^{1/3}&  n=2\\ 
(2/5^{1/4}) (M/\md)^{1/4}& n=3\\ 
(5\pi)^{1/5}  (M/\md)^{1/5}& n=4 \\ 
2\left(3\pi/7\right)^{1/6}(M/\md)^{1/6} & n=5\\ 
(105 \pi^2/2)^{1/7} (M/\md)^{1/7} & n=6 \\
2(2\pi)^{1/4}/3^{1/8}  (M/\md)^{1/8} \approx 2.76  \mt^{1/8} & n=7.
\end{matrix}\right. 
\end{eqnarray}

\begin{eqnarray}
\frac{4\pi T_h}{\md}&=&
\left\{\begin{matrix}
\sqrt{6\pi} (M/\md)^{-1/2} \approx 4.34 \mt^{-1/2} & n=1\\ 
6^{2/3}(M/\md)^{-1/3}&  n=2\\ 
2 \cdot 5^{1/4} (M/\md)^{-1/4}& n=3\\ 
(5^4/\pi)^{1/5}  (M/\md)^{-1/5}& n=4 \\ 
(3^5\cdot 7/\pi)^{1/6}(M/\md)^{-1/6} & n=5\\ 
(2\cdot 7^6/3\cdot 5\cdot \pi^2)^{1/7} (M/\md)^{-1/7} & n=6 \\
(2^6\cdot 3/\pi^2)^{1/8}  (M/\md)^{-1/8}\approx 2.90 \mt^{-1/8}& n=7.
\end{matrix}\right. 
\end{eqnarray}

\begin{eqnarray}
S_h &=&
\left\{\begin{matrix}
\sqrt{32\pi/27} (M/\md)^{3/2} \approx 1.92 \mt^{3/2} & n=1\\ 
(3/4)^{2/3}\pi (M/\md)^{4/3}&  n=2\\ 
8\pi/5^{5/4}(M/\md)^{5/4}& n=3\\ 
 2(5\pi^6)^{1/5}/3 (M/\md)^{6/5}& n=4 \\ 
8(3\cdot \pi^7/7^7)^{1/6}(M/\md)^{7/6} & n=5\\ 
(105\cdot \pi^9/2^8)^{1/7}  (M/\md)^{8/7} & n=6 \\
8 \cdot (2\pi^5)^{1/4}/3^{17/8}  (M/\md)^{9/8}\approx 3.85 \mt^{9/8}& n=7.
\end{matrix}\right.
\end{eqnarray}

\newpage


\section{Physics in trans-Planckian energy domain\label{sec:energy domain}}

\begin{quote}
``The scattering process of two pointlike particles at CM energies in the order of Planck units or beyond is very well calculable using known laws of physics because graviton exchange dominates over all other interaction processes. At energies much higher than the Planck mass,  black hole production sets in, accompanied by coherent emission of real gravitons.\\
--G. 't Hooft (1987)"
\end{quote}

As 't Hooft provided these preliminary arguments in 1987 \cite{'tHooft, Dray}, which have been clarified by many subsequent studies, gravity becomes dominant in the trans-Planckian domain and  the `known laws of physics', i.e., a combination of quantum field theory and general relativity provides a good description of certain features of trans-Planckian physics. In this section, we review the scattering problem in the trans-Planckian energy domain, which is described by the known physics in the 't Hooft sense. We use the conventional terminology in particle physics $\sqrt{s}$ instead of $E_{\text{c.m.}}$ for the scattering energy in the center of the collision frame.  Relatively little is known in the Planck domain ($\sqrt{s}\sim M_D$), but a large part of the trans-Planckian domain ($\sqrt{s}\gg M_D$) and sub-Planckian domain ($\sqrt{s} \ll M_D$) can be understood in terms of known physics, i.e., perturbative quantum field theory and Einstein gravity in higher dimensions.

\subsection{The limits of the effective field theory approach}

In the effective field theory framework, an action for describing physics below a high scale or a cut-off scale $\Lambda$ is composed of  a set of quantum field operators, which may be or may not be suppressed by the $\Lambda$ scale depending on their dimensions. An operator ${\cal O}_{4+d}$ of the mass dimension $(4+d)$ is suppressed by $\Lambda^d$ as the Lagrangian density ${\cal L}$  has the mass dimension $4$ or the number of spacetime dimensions. An operator that  is suppressed by $\Lambda^{d>0}$ is regarded as unimportant or irrelevant at low energy, $q^2 \ll \Lambda^2$, and its effect is regarded as a small correction to the effect not suppressed by $\Lambda$. The generic form of the Lagrangian density in flat spacetime with the Lorentzian metric $\eta_{\mu\nu}=\text{diag}(1,-1,-1,-1)$  is given by the following schematic expression:
\begin{eqnarray}
{\cal L}_{\text{eff}} ={\cal O}_4 + {\cal O}_5/\Lambda + {\cal O}_6/\Lambda^2  +\cdots.
\end{eqnarray}
This approach clearly fails when one considers a physical process beyond the cut-off scale $\Lambda$. In this ``trans-cut-off domain" all the higher order terms become equally important and their effects can no longer  be regarded as small corrections  to the leading order result. A higher-scale new physics is needed 
before this effective field theory fails. A good historical example is Fermi's four fermi theory of weak interactions, which failed on high scales but the electroweak theory of $SU(2)\times U(1)$ takes over the role of the new gauge bosons $W, Z$ and $\gamma$ (photon). 

 Now we can consider an effective field theory for matter fields coupled with gravity.\footnote{A variation of  the Einstein-Hilbert action is given as $\delta S_{\text{EH}}= -\frac{1}{16\pi G} \int d^4 x \sqrt{g} \left[R_{\mu\nu}-\frac{1}{2}g_{\mu\nu}R\right]\delta g^{\mu\nu}$. The energy-momentum tensor of the matter Lagrangian is {\it defined} by $\delta S_m = \frac{1}{2}\int d^4 x \sqrt{g} T_{\mu\nu} \delta g^{\mu\nu}$ so that $\delta S_{\text{EH}}+\delta S_m=0$ leads the Einstein equation $R_{\mu\nu}-\frac{1}{2}g_{\mu\nu}R -8\pi G T_{\mu\nu}=0$.} A mass scale is set by the Planck scale,  $\Lambda\sim M_4$, so that the  effective Lagrangian for matter fields $S_m$  is given in terms of operators, which may or may not be suppressed by $M_4$:%
\begin{eqnarray}
S=S_{\text{EH}} + \int d^4 x \sqrt{g} ({\cal O}_4 + {\cal O}_5/M_4 + {\cal O}_6/M_4^2  +\cdots ).
\end{eqnarray}
 When one considers physical processes in a low energy regime with small energy transfers, $q^2 \ll M_4^2$, higher order terms ${\cal O}_{>4}$ are irrelevant. On the other hand, when $q^2 \sim M_4^2$, all higher order terms become non-negligible thus this perturbative description of the effective theory would not work. Indeed, if the relevant energy is well below $M_4$, gravity is weak (the weakest force of all known fundamental forces). One can consider the gravitational interaction a small perturbation to the known non-gravitational interactions.  When the energy becomes high and reaches the Planck energy, the gravitational interaction becomes dominant over all other forces  as the gravity directly couples to the energy-momentum tensor differently from gauge interactions. 
 However, one should note that  a large momentum transfer can be carried by many soft exchanged gravitons, in which case the higher-dimensional operators appear not to play a role \cite{Giddings:2010}.

In $D=4+n$ dimensions, a higher order term in addition to the Einstein-Hilbert actions is written as%
\begin{eqnarray}
S=\int d^D x \sqrt{G_D} \frac{M_D^{D-2}}{2} \left(R + {\cal O}\left(R^2 /M_D^2\right) + \cdots \right).
\end{eqnarray}
The actual form of the second order term is suggested in string theory, e.g., $a R^2 + b R^{\mu\nu}R_{\mu\nu} + c R^{\mu\nu\rho\sigma}R_{\mu\nu\rho\sigma}$ with $(a,b,c)= (1,-4,1)$ or the Gauss-Bonnet term; there is  no ghost in $4D$. The second and higher order terms can be neglected when the curvature (=Ricci scalar) is `small' or $R/M_D^2 \ll 1$ where we get the conventional Einstein-Hilbert action.


\subsection{The trans-Planckian domain: $\sqrt{s} \gg  M_D$}

First let us consider  relevant length scales for the  trans-Planckian scattering of the two particles. They are the Planck length ($\ell_D$), the de Broglie length of the colliding particles ($\ell_{\rm dB}$), the Schwarzschild-Tangherlini black hole length ($r_s$) corresponding to the CM energy and the impact parameter of the scattering in terms of angular momentum \cite{Giudice:2001}:
\begin{eqnarray}
&&\ell_D=\left(\frac{G_D\hbar}{c^3}\right)^{\tfrac{1}{D-2}},\\
&&\ell_{\rm dB}=\frac{4\pi \hbar c}{\sqrt{s}},\\
&&r_s=C_D \left(\frac{G_D\sqrt{s}}{c^4}\right)^{\tfrac{1}{D-3}},\\
&&b=2 \frac{J}{\sqrt{s}},
\end{eqnarray}
where
$C_D=\left(\frac{16\pi}{(D-2)\Omega_{D-2}}\right)^{\tfrac{1}{D-3}}$ and $\Omega_{D-2}$ is the area of the unit sphere, $S^{D-2}$.

It is interesting to notice in the limit with $\hbar \to 0$ and $G_D, \sqrt{s}$ fixed, 
\begin{eqnarray}
\ell_D\to 0, \,\ell_{\rm dB} \to 0,\,r_s \to {\rm finite},
\end{eqnarray}
which means that {\it semi-classical physics is responsible for  trans-Planckian scattering} and the characteristic
length scale for the dynamics is $r_s$. For instance, when the impact parameter of the collision is $b$, one can estimate the small scattering angle $\theta\sim (r_s/b)^{n+1}$ as is confirmed by the Eikonal approximation in the trans-Planckian regime (see earlier work in string theory \cite{Amati:1987a, Amati:1987b, Amati:1988, Amati:1990, Amati:1993, Damour:1999}
and also see the relatively recent work \cite{Giudice:2001,Giddings:2009, Giddings:2010, Stirling:2011}). As $b$ approaches $r_s$,  non-linear gravitational attraction becomes important and at  around $b\sim r_s$, black hole formation is expected.

A quantitative definition of the trans-Planckian domain ($\cal A$) for a black hole  is given in terms of the Kretschmann invariant ${\cal K}\equiv \left|R^{LMNP}R_{LMNP}\right|^2/\left((D-2)\sqrt{(D-1)(D-3)} E_P^2\right)$ in $D$ dimensions by Okawa et al. using a convenient energy scale $E_P \equiv \left((D-2)\Omega_{D-2}/16\pi G_D\right)^{1/(D-1)}$ 
\cite{Okawa:2011, Okawa:2011b} as follows:
\begin{eqnarray}
\text{trans-Planckian domain:} \quad {\cal A} = \{{\cal A}: \underset{\cal A}{\rm Inf}\,{\cal  K} \geq 1 \},
\end{eqnarray}
where the inequality is saturated for a black hole with a horizon of the size of the corresponding Compton wave length of the same mass. In the regime ${\cal A'}$ with $\underset{\cal A'}{\rm Inf}\,{\cal  K} \ll 1$, we can solely trust classical approximation over which quantum effects are suppressed as ${\cal O}({\cal K}) \ll1$.

\subsubsection{Parton level collision}

When we consider the process of a two-particle collision in asymptotically flat spacetime we can define a reference frame for the center of mass (or CM frame) in which the total sum of spatial momenta vanishes at the asymptotic limit. The CM energy is also obtained by summing the energies  in the same reference frame. If the CM energy is significantly larger than the Planck energy, we call the collision  trans-Planckian or super-Planckian. If the particles participating in the collision are composite or non-elementary, the actual CM energy contributing to the collision should be obtained for `partons', which could be quarks and gluons in hadron colliders such as the LHC. This CM energy of partons is called the parton level CM energy, which often is much greater than the sum of the rest mass energies of partons due to a large kinetic energy contribution of the partons. On the other hand, the parton level CM energy is smaller than the total collision energy because other particles (spectators) do not contribute to the collision,  but still carry energy away. If two partons have the same mass, the parton level CM energy is 
\begin{eqnarray}
E_{\rm CM} = \frac{2 m}{\sqrt{1-v^2}} \equiv 2 m \gamma(v),
\end{eqnarray}
where $v<1$ is the velocity of each particle in the CM frame. As defined above, the collision is called trans-Planckian when $E_{\rm CM}\gg M_D$ in $D$-dimensions or
\begin{eqnarray}
 2 m \gamma(v) \gg M_D \,\,\,\text{trans-Planckian},
\end{eqnarray}
thus for partons of mass $ m\ll M_D$, a large $\gamma$ factor is required.  For instance,  light quarks (up-quark, down-quark) and gluons are partons in a proton, and each parton carries a part of the proton energy ($E_{\rm parton}= x E_{\rm proton} \sim \gamma  (x m_{\rm proton}), 0\leq x \leq 1$) but still can be larger than its rest mass energy.  The required $\gamma$ factor is estimated as
\begin{eqnarray}
\gamma(v) \gg \frac{M_D}{\sqrt{x_1 x_2 m_{\rm proton}^2} }\sim \frac{1}{\sqrt{x_1 x_2} } \frac{M_D}{m_{\rm proton}} \sim \frac{1}{\sqrt{x_1 x_2}} (M_D/{\rm TeV}) \times 10^3,
\label{eq:ballpark}
\end{eqnarray}
which defines the ballpark parameter range for the LHC study. On the other hand, when we consider `heavy' objects, especially colliding black holes, their masses could be assumed to be much larger than $M_D$ so that we may still reach the trans-Planckian domain with a rather small $\gamma\sim {\cal O}(1)$. This domain is nothing but the classical physics domain.

\subsubsection{Black hole formation}

\begin{figure}[h]
\begin{center}
        \includegraphics[width=.3\textwidth]{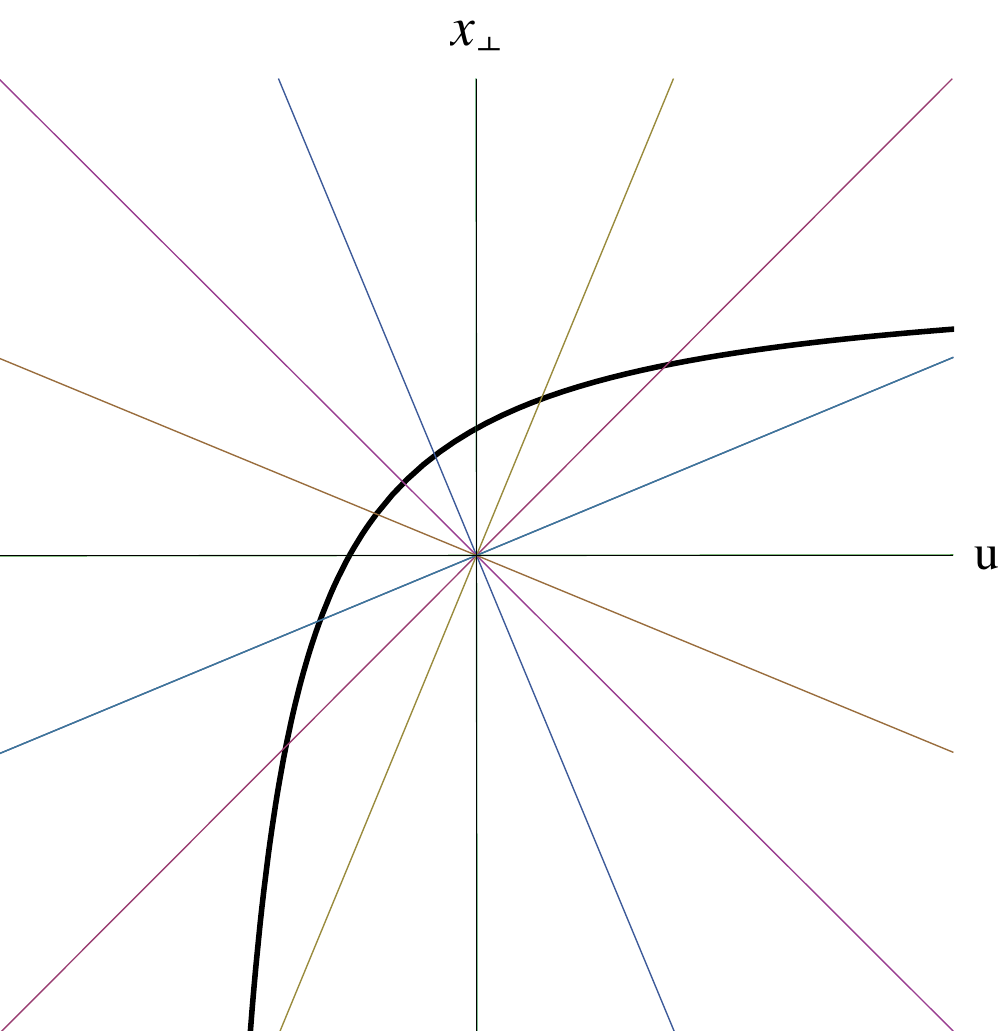}
                \includegraphics[width=.3\textwidth]{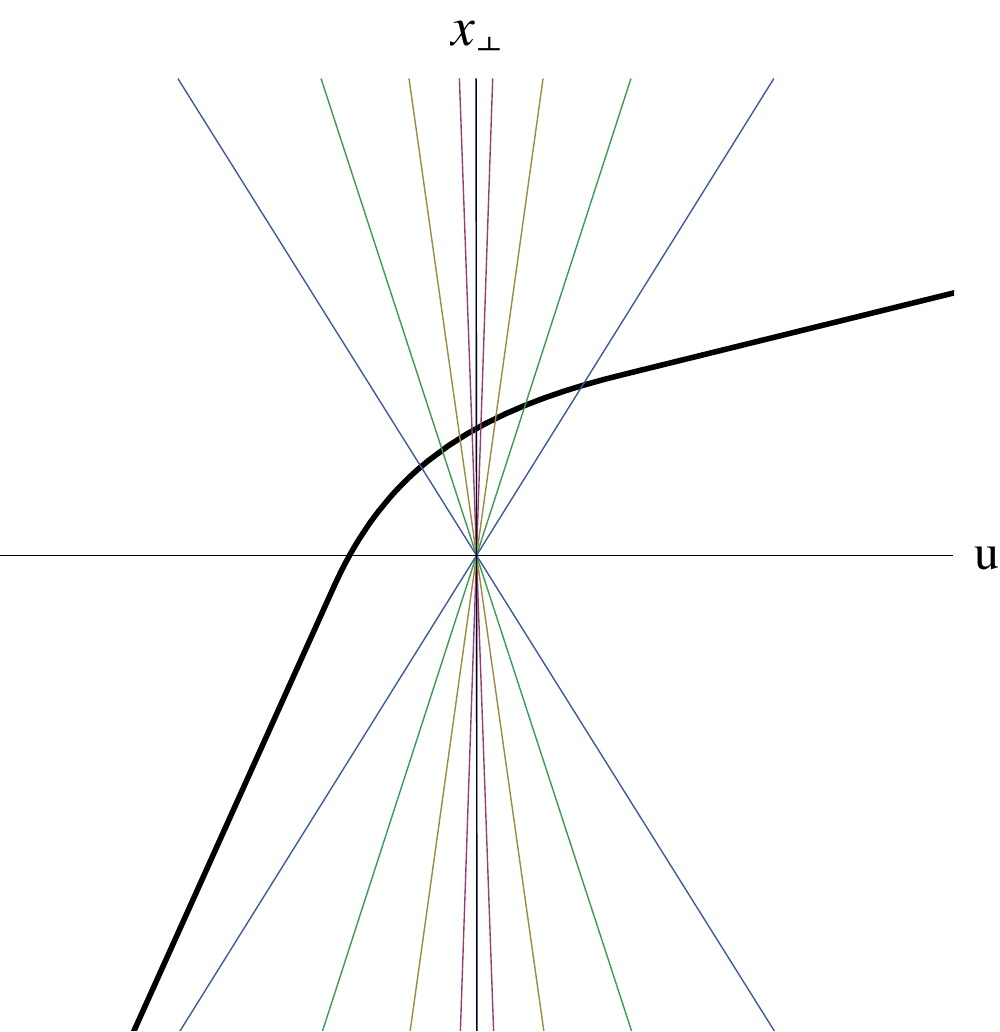}
        \includegraphics[width=.3\textwidth]{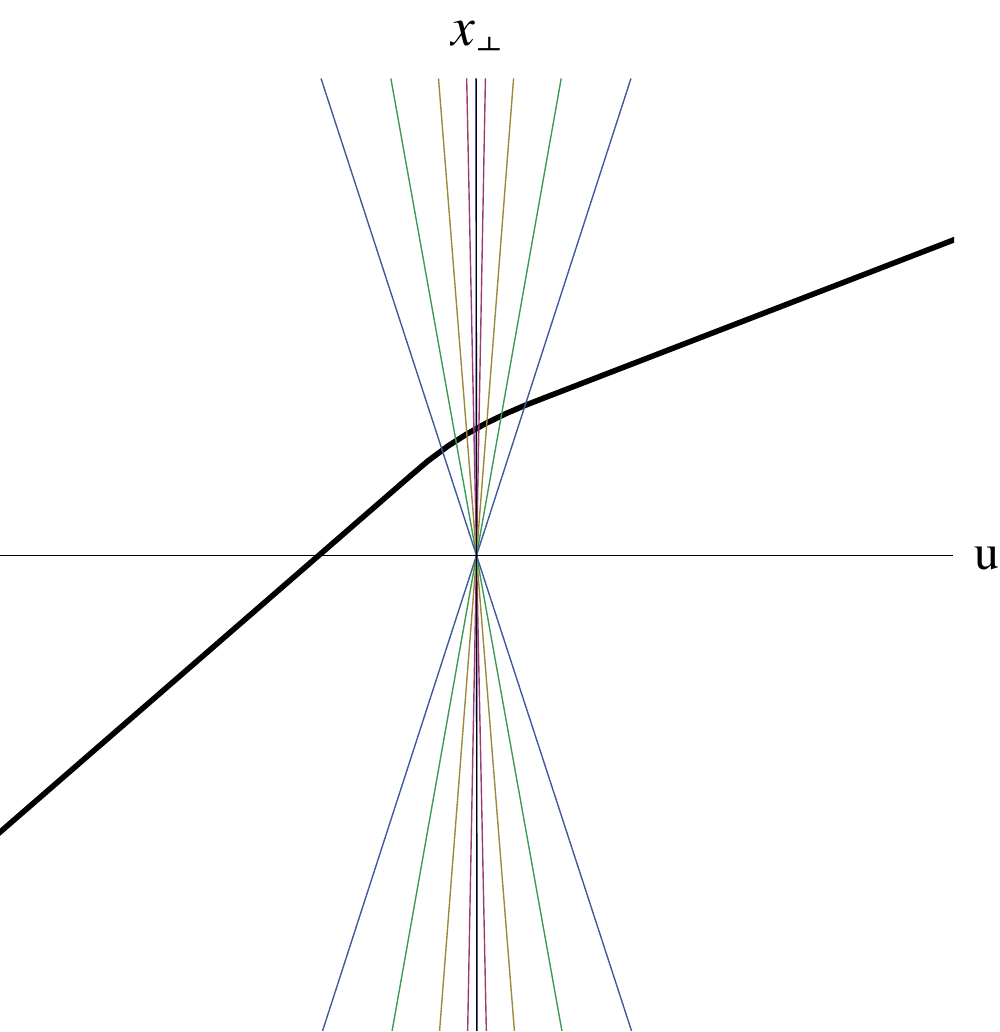}
        \end{center}
  \caption{Scattering trajectories of a slowly moving particle by a stationary particle, a slowly moving particle and an ultra relativistic particle, respectively (from left to right). The black thick lines are the trajectories. The colorful thin lines indicate the gravitational field associated with the target particle at $u=z-v t=0$. }
\label{fig:collisions}
\end{figure}

 A particularly interesting phenomenon is black hole formation in trans-Planckian collision with a sufficiently small impact parameter.  The black hole formation in particle collision can be easily understood by considering two infinitely boosted particles (take $m\to 0$, $v\to 1$  and $\sqrt{s}={\rm fixed}$). When a particle is highly boosted, the field line of gravity of the boosted particle is well confined on the transverse plane of the momentum. This highly boosted gravitational field is described by a gravitational shock wave identified by Aichelburg and Sexl \cite{Aichelburg:1971} and the superimposed solution by Dray and 't Hooft in 1985 \cite{Dray} for $D=4$ and Ferrai et al. for $D>4$  in 1988  \cite{Ferrari:1988}. (see \cite{Kaloper:2007} for an intuitive exposition on black hole formation in high energy collisions). In Fig. \ref{fig:collisions}, for illustration, we plotted the gravitational field lines for a rest particle, a boosted particle and  a highly boosted particle, respectively, from the left to the right. The solid black curves describe the orbit of a slowly moving particle in the presence of the boosted gravitational field.  A slowly moving particle can `feel' the gravitational attraction at a distance if the source particle is at rest (figure on the left). However, if the source is highly boosted (the right-most figure), because the field is confined on the transverse plane, the slowly moving particle feels the gravitational attraction right at the moment of the encounter with the gravitational shock wave not before and after. Thus the slowly moving particle behaves just like a free particle before and after the encounter (see the third figure in Fig. \ref{fig:collisions}). On the transverse plane, the gravitational field satisfies the Poisson equation 
 \begin{eqnarray}
 \nabla_i^2 \Phi(x^i) = -16\pi G_D \mu \delta^{D-2}(x^i),
 \label{eq:Poisson}
 \end{eqnarray}
  where $x^i$ denotes transverse coordinates. The solution  is easily found:
\begin{eqnarray}
\Phi =\left\{\begin{matrix}
-8\pi G \mu \log \frac{\sqrt{x^i x_i}}{L} && ,D=4\\ 
16\pi G_D\mu /(\Omega_{D-3}(D-4) (x^ix_i)^{(D-2)/2}) && ,D=4+n.
\end{matrix}\right.
\end{eqnarray}

The metric is now given as
  \begin{eqnarray}
  ds^2 = du dv -dx^{i2} -\Phi(x^i) \delta(u) du^2,
  \end{eqnarray}where the light cone coordinates are $u=t-z$, $v=t+z$ and  $r$ and $\phi$ are the coordinates on the $(D-2)$-plane. The metric is flat except for $u=0$.\footnote{Adopting a convenient unit of length $r_0 =(8\pi G_D\mu/\Omega_{D-3})^{1/(D-3)}$ and the continuous coordinates, 
\begin{eqnarray}
&&u \to u,\\
&&v \to \left\{\begin{matrix}
v-2 \log r\theta(u)+ u \theta(u) & ,D=4 {}\\ 
v+2\theta(u)/(D-4)r^{D-4} + u\theta(u) r^{6-2D} & ,D=4+n.
\end{matrix}\right.\\
&&r \to r\left(1-u \theta(u) r^{2-D}\right),\\
&&\phi_i \to \phi_i,
\end{eqnarray}
the metric is given
\begin{eqnarray}
ds^2 = du dv -\left(1+(D-3) \frac{u}{r^{D-2}}\theta(u)\right)^2 dr^2 +r^2 \left(1-\frac{u}{r^{D-2}}\theta(u) \right)^2 d\Omega_{D-3}^2,
\end{eqnarray}
where $\theta(u)$ is the Heaviside step function, which vanishes when $u<0$.}

\begin{figure}[h]
\begin{center}
    \includegraphics[width=.55 \textwidth]{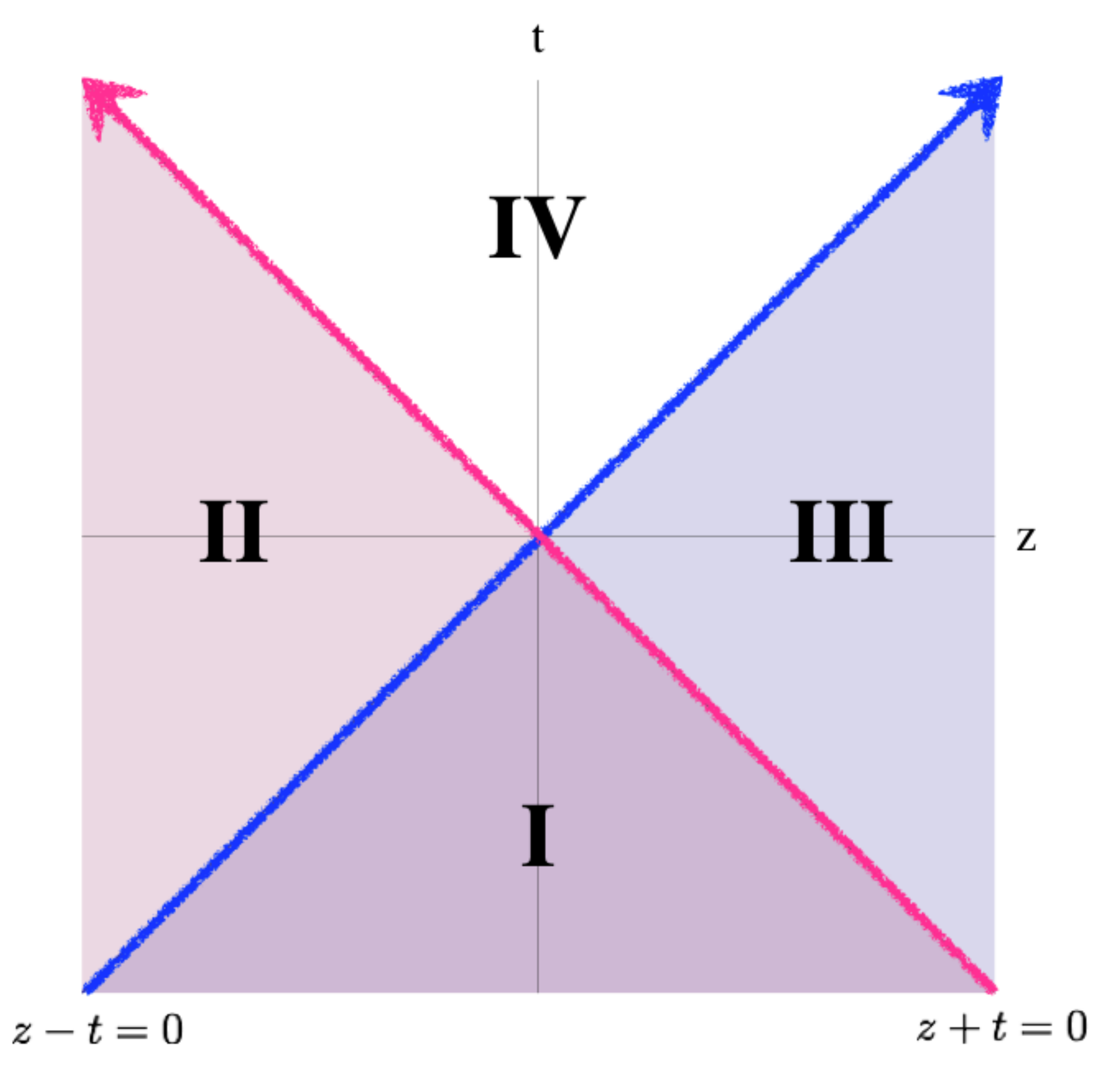}
  \caption{ Spacetime diagram denoting two shock wave collisions. The left parton moves along $z-t=0$ with the speed of light and right partons move in opposite direction and collide at the center. Regions I, II and III are described by the combined Aichelburg-Sexl (AS) metric and are thus completely determined. Right outside of the Yoshino-Rychkov (YR) slice (outside of IV) the AH formation is numerically determined, and thus the bound for $b_{\rm Max}$ can be calculated. The future light cone (IV) is not determined by the AS metric such that $b_{\rm Max}$ on the YR slice is the best lower bound on $b_{\rm Max}$. }
\label{fig:lightcone}
\end{center}
\end{figure}

In Fig. \ref{fig:lightcone},  we plotted a spacetime diagram of colliding shock waves. A shock wave traveling along $v=t+z=0$ in the $-z$ direction will not be influenced until the shock waves collide so that the superposition of the two shock wave solutions gives the exact geometry in regions I, II and III outside the future light cone  (region IV) of the collision of the shocks.

Having the explicit combined shocks, one can check the appearance of the apparent horizon (AH, a marginally trapped surface ${\cal S}$, the outer null normals of which have zero convergence) in regions I, II and III for various values of impact parameter $b$. The best bound on the maximum  impact parameter, $b_{\rm max}$, can be obtained on the Yoshino-Rychkov slice (outside region IV).  If the apparent horizon (AH) exists, it is necessarily inside of the event horizon so the presence of AH guarantees the formation of an event horizon (EH) or a black hole according to the area theorem. Penrose ($D=4, b=0$) \cite{Penrose:1974}, Eardley-Giddings ($D\geq 4, b\geq 0$, a trapped-surface construction) \cite{formation1}, Yoshino-Nambu ($D>4, b\geq 0$, numerical solutions) \cite{formation2} and Yoshino-Rychkov ($D>4, b\geq 0$, refined numerical solutions) \cite{formation3}  studied the formation of AH in shock collisions for various cases.   From $b_{\rm max}$ one can directly calculate the lower bound of the cross-section of black hole formation ($\sigma_{BH} >\sigma_{AH}= \pi b_{\rm max}^2$).

The maximum value of the impact parameter can also be obtained under the assumptions of the Hoop conjecture, which was originally suggested by Thorne in 1972 \cite{Thorne:1972}. In four dimension, the Hoop conjecture states that a black hole with horizons forms when and only when a mass $M$ is  compacted into a region with a  circumference in every direction, ${\cal C}\lsim 4\pi G_4 M$, where $G_4$ is the Newton's constant in four-dimensional general relativity. For higher dimensional black holes, Ida and Nakao \cite{Ida:2002} suggested an isoperimetric inequality $V_{D-3} \lsim G_D M$, where $V_{D-3}$ denotes the volume of the typical closed $(D-3)$-section of the horizon (hyperhoop), which is apparently different from a naive expectation ${\cal C}\lsim (GM)^{1/(D-2)}$.

\begin{figure}[h]
\begin{center}
        \includegraphics[width=.65\textwidth]{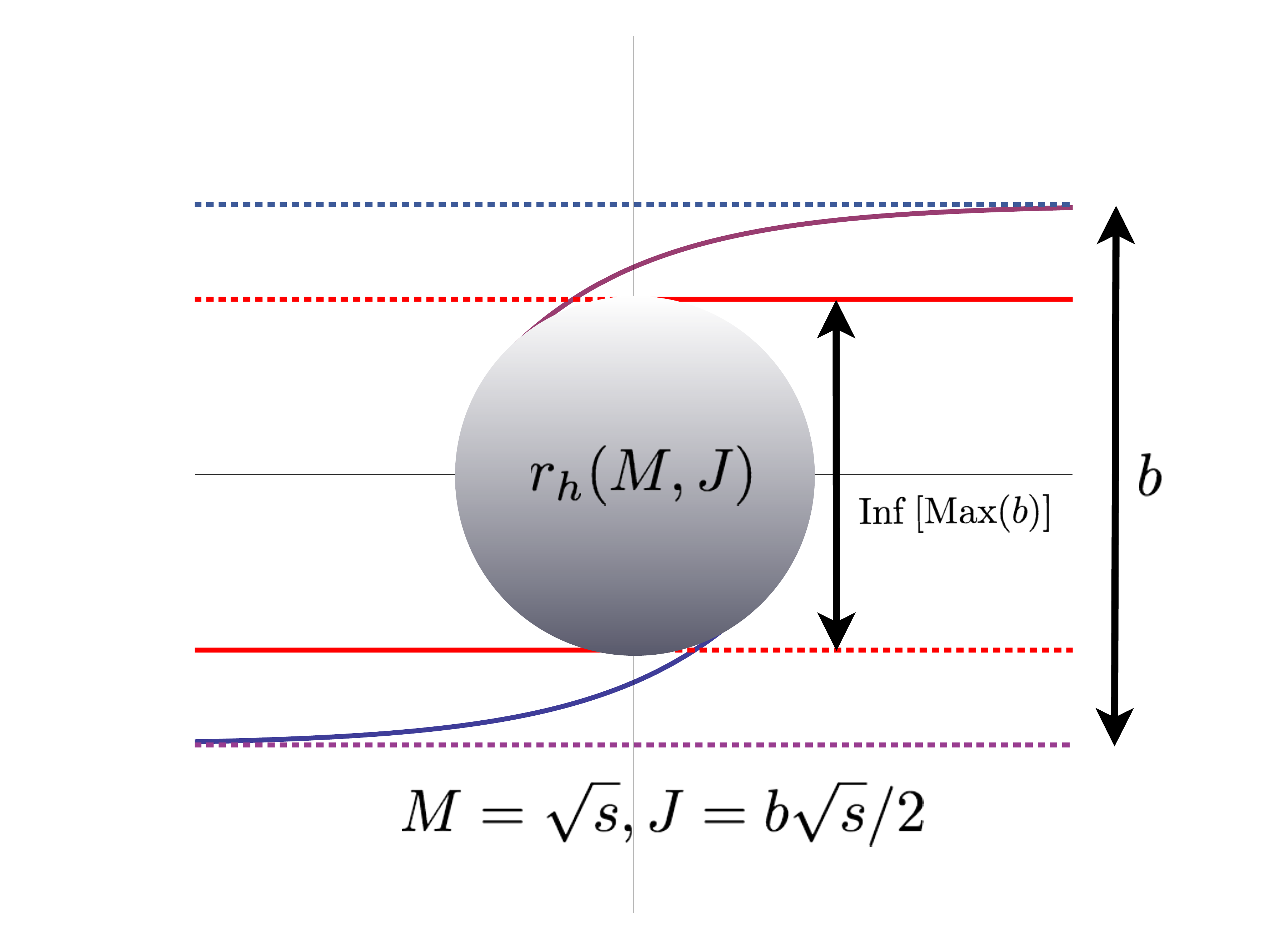}
                       \end{center}
  \caption{Schematic view of black hole formation in particle collisions. The Hoop conjecture provides a criterion for a black hole to form:
  $b\leq 2 r_h(M,J)$. Here $b_{\rm Max}$ saturating the inequality should be understood as the  infimum of the maximum value for the allowed impact parameter for black hole formation.}
\label{fig:scheme}
\end{figure}

For colliding particles, Park(2001) \cite{Park:2001} and Ida-Oda-Park(2002) \cite{IOP1} generalized the original Hoop conjecture by taking into account the angular momentum as follows (also see work by Yoo et al.(2009) \cite{Yoo:2009}). 
Let us consider a collision of two massless particles with a finite impact parameter $b$ and CM energy $\sqrt{s}$ such that each particle has an initial energy $\sqrt{s}/2$ in the CM frame. The initial angular momentum is also determined by $J_i = b \sqrt{s}/2$ in the CM frame. Suppose that a black hole forms when the two particles come into a region surrounded by a Hoop with a radius of the event horizon of the mass $\sqrt{s}$ and angular momentum $J$. Then the infimum of the maximum impact parameter for a black hole to be formed during the collision is
\begin{eqnarray}
{\rm Inf}\left[{\rm Max}(b (\sqrt{s})\right]= 2 r_h(\sqrt{s},J=b\sqrt{s}/2). 
\label{eq:Hoop}
\end{eqnarray}
Hereafter, for the brevity of notation, we simply denote $b_{\rm Max}={\rm Inf}\left[{\rm Max}(b (\sqrt{s})\right]$.

Using the relations in Eq. \eqref{eq:MJ} and also in Eq. \eqref{eq:rh}, one can easily found the maximum value of the impact parameter is set saturating the the inequality $b<b_{\rm Max}$ \cite{Park:2001, IOP1}:
\begin{eqnarray}
b_{\rm Max}(\sqrt{s}) = \frac{2}{\left(1+\left(\tfrac{D-2}{2}\right)^2\right)^{\tfrac{1}{D-3}}} r_s(\sqrt{s}),
\label{eq:bmax}
\end{eqnarray}
which can be compared with the numerically obtained values by Yoshino and Rychkov.
When $b = b_{\rm Max}$, the rotation parameter $a_*$  takes the maximal value $(a_*)_{\rm Max} = (n + 2)/2$ and the corresponding angular momentum is $J_{\rm Max}=b_{\rm Max}\sqrt{s}/2$. Assuming the production rate is unity when $b \leq b_{\rm Max}$, the production cross section of black hole can be expressed as
\begin{eqnarray}
\sigma (\sqrt{s}) \geq  \pi  b_{\rm Max}^2 (\sqrt{s}).
\end{eqnarray}

In Table \ref{table:bmax} we summarize the results obtained by Yoshino-Rychkov and Ida-Oda-Park and compare them with the naive value obtained by dimensional analysis. One should notice that the analytic calculation in Ida-Oda-Park assumes (1) no attraction of the geodesics before the collision, (2) no radiation of energy during the collision; both of these assumptions {\it reduce} the estimated maximum value of $b$ or {\it enlarge} the ratio $\sigma/\pi b_{\rm max}^2$. In this sense, the results by reported by Yoshino-Rychkov are remarkably consistent with the results described by Ida-Oda-Park.

It is possible to study the perturbation of the geometry by superimposing two equal Aichelburg-Sexl shock waves traveling in opposite directions  in $D=4+n$ dimensions.  D'Eath and Payne (1992) studied black hole-black hole collisions in $D=4$ and made a crude estimate of the efficiency of gravitational wave generation, about 16.4\%  \cite{D'Eath:1992a, D'Eath:1992b, D'Eath:1992c}. More recently, it was found that about $25.0\%$ ($D=4$) and $41.1\%$ ($D=10$) of the initial energy is radiated as gravitational waves in head-on collisions \cite{Herdeiro:2011}. It is interesting to note that the percentage of the emission {\it increases} monotonically with respect to the larger number of dimensions. This result is understandable as the number of degrees of freedom in gravitational waves in higher dimensions, is {\it bigger} than the one in lower dimensions. 

\vspace{.5cm}
\begin{table}[h]
\begin{center}
\begin{tabular}{c|cccccccc}
 $D$ & $4$ & $5$ &$6$ & $7$ & $8$ & $9$ & $10$ & $11$ \\
  \hline
  \hline
$\sigma_{\rm YR}/ \sigma_{\rm IOP}$ & $0.71$ & $1.25$ &$1.57$ & $1.69$ & $1.73$ & $1.75$ & $1.73$ & $1.72$ \\
$\sigma_{\rm YR}/\pi r_s(M)^2$ & $0.71$ & $1.54$ &$2.15$ & $2.52$ & $2.77$ & $2.95$ & $3.09$ & $3.20$ \\
\end{tabular}
\caption{The value of $\sigma_{\rm YM}$ obtained by Yoshino-Rychkov \cite{formation3} in comparison with the value analytically obtained by Hoop conjecture, $\sigma_{\rm IOP}=\pi b_{\rm max}^2$, in Ida-Oda-Park \cite{IOP1} and also a naive dimensional analysis $\sigma\approx \pi r_s^2$. }
\end{center}
\label{table:bmax}
\end{table}

%
%
\subsubsection{Numerical relativity on high velocity collisions \label{sec:numerical}}

Only a  limited amount of information could be obtained by Aichelburg-Sexl solutions. One may guess that a black hole forms when $b<b_{\rm max}$ such that a differential cross-section of black hole formation is simply given by a geometric cross section:
\begin{eqnarray}
\left[\frac{d\sigma}{d b}\right]_{\rm geo} = \frac{\pi b}{2}, 
\end{eqnarray}
or $d \sigma/d J \propto J$ since $J \propto b$ \cite{IOP1}, which gives the conventional result $\sigma =\int_0^{b_{\rm Max}} d b \frac{d\sigma}{d b} = \pi b_{\rm Max}^2$. However, the actual formation is dynamical and the process is violent. Numerical relativity is one of the tools used to study this dynamical situation.

For the LHC study, the following quantities are absolutely important  for a given set of initial data $(\sqrt{s},b)$:
\begin{itemize}
\item How much energy will be lost during the high energy collision? (Graviational radiation during the balding phase): $E_{\rm final}$
\item How large of an angular momentum will be lost during the collision? (Angular momentum extraction during the balding phase): $J_{\rm final}$
\item Production cross-section for a given $(E,J)_{\rm final}$: $\frac{d\sigma}{dE dJ}$
\end{itemize}

Indeed we have observed impressive progress in this direction during the last five years or so. The ballpark value for $\gamma(v) \gsim 10^3$ (see Eq. \eqref{eq:ballpark}), which may not be reachable within a short foreseeable time scale but still the recent progress could provide valuable information.

\begin{itemize}
\item Higher dimensional numerical relativity formalism:  Sorkin (2009) \cite{Sorkin:2009}, Shibata-Yoshino (2009) \cite{Shibata:2009}, Zilhao et al. (2010) \cite{Zilhao:2010}, Lehner-Pretorius (2010) \cite{Lehner:2010}.
\item Numerical simulation for high velocity collision: Yoshino et al. \cite{Yoshino:2005aa}, Sperhake et al. (2008) \cite{Sperhake:2008},  Shibata et al. (2008) \cite{Shibata:2008}, Sperhake et al. (2009) \cite{Sperhake:2009}, Witek et al. (2010) \cite{Witek:2010},  Okawa et al. (2011) \cite{Okawa:2011}.
\item Bar-mode instability: Shibata and Yoshino (2010)  \cite{Shibata:2010}.
\end{itemize}

\subsubsection{Planck domain: $\sqrt{s}\sim M_D$}
Relatively less is known for this domain. In this domain, the de Broglie wave length is comparable to the Planck length so that a geometrical notion of length does not have a concrete physical sense:
\begin{eqnarray}
\lambda_{\rm de\, Broglie} \sim \ell_{\rm Planck}.
\end{eqnarray}
In string theory, if the string scale is below the Planck scale, $M_S \sim TeV \lsim M_D$  ``stringy effects'' could show up in the Planck domain.
In principle, string-black hole correspondence, stringy excitation modes of the order of $M_S$, string balls as a highly excited string state, and the  a characteristic high energy behavior of string scattering cross-sections could be included in the list of ``stringy effects" and some  quantitative predictions have been attempted  (see Cullen, Perelstein, Peskin (2000) Horowitz(1996) \cite{Horowitz:1996}, \cite{Cullen:2000}, Dimopoulos (2001) \cite{Dimopoulos:2001st}, Oda-Okada (2001) \cite{Oda:2001}, Anchordoqui et al.(2009) \cite{Anchordoqui:2009}, Nayak(2009) \cite{Nayak:2009}).

\subsection{A phase diagram}

\begin{figure}[h]
\begin{center}
        \includegraphics[width=.85\textwidth]{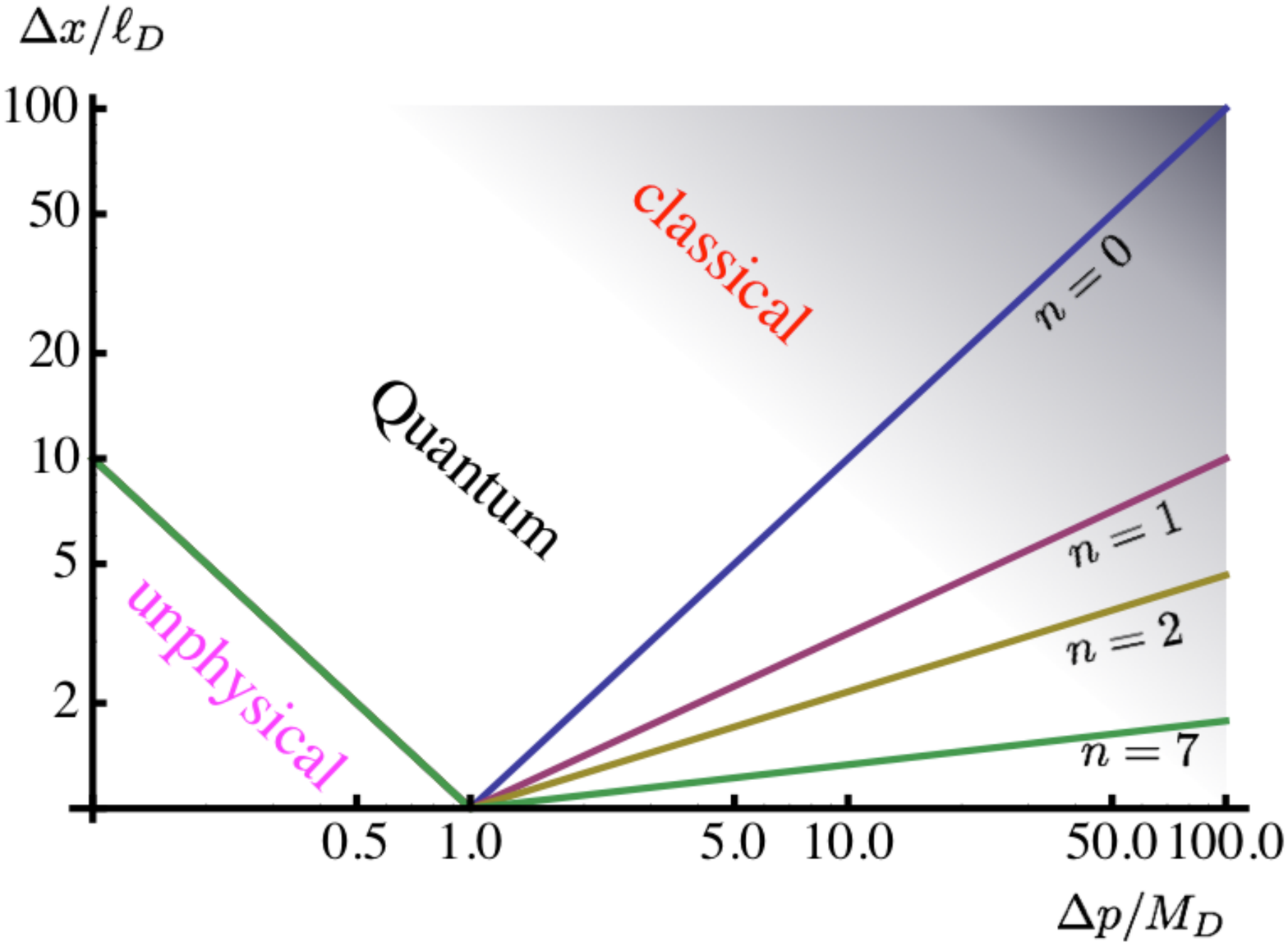}
        \end{center}
  \caption{A phase diagram describing the minimum distance scale to be probed with a given energy (momentum transfer). Below the Planck energy $M_D$, the minimum distance is determined by the uncertainty 
  relation $\Delta x \geq \hbar/\Delta p$. Above a certain energy, presumably Planck energy, the information
  behind the event horizon is hidden and the minimum distance is essentially determined by the size of the classical Schwarzschild 
  radius $r_\text{Sch}\sim E^{1/(n+1)}$  for the given energy.  The domain with a large distance $\Delta x \gg \hbar/\Delta p$ corresponds to a classical domain where quantum corrections are small. $\Delta x < \hbar/\Delta p$ is unphysical.}
\label{fig:uncertainty}
\end{figure}

 A suggested `phase diagram' in Fig. \ref{fig:uncertainty} explains the energy-distance relation (see  Park (2008) \cite{Park:2008}, Giddings (2010) \cite{Giddings:2010}, Dvali et al. (2011) \cite{Dvali:2011} ). In the figure, the horizontal axis is for the amount of available energy in physical processes such as the scattering of two particles. The vertical axis represents the smallest distance scale that can be probed by the corresponding energy. In a domain well below the Planck energy ($M_D$), the distance scale is limited by the uncertainty principle or $\Delta x \gsim 1/\Delta p$. As the energy increases, the quantum gravitational effects become more and more relevant and finally dominate over all other interactions. On the other hand, once the energy surpasses the Planck domain, a new window of black hole production would be available and then the smaller scale physics ($\ell \ll \ell_D$) would be hidden behind the event horizon corresponding to the energy such that finally the smallest distance scale is essentially given by the size of the Schwarzschild radius $r_s\sim G s$. In the presence of extra dimensions, the energy dependence of the Schwarzschild radius is $r_s\propto M^{1/(D-3)}$ in $D$-dimensions (more discussion is in the next section) so that $\Delta x \gsim \Delta p^{1/(D-3)}$.

\newpage


\section{Black holes at the LHC \label{sec:LHC}}

The primary goal of the LHC is to find the Higgs boson to confirm the standard model of particle physics.  The other goal, which is equally important,  is to search for signatures from new physics around $1$ TeV. Low scale gravity models are subject to be tested by the LHC.  The various quantum gravitational effects in particle interactions as well as the black hole formation are evident  in these models. Currently, the LHC accumulates data with a high integrated luminosity and $\sqrt{s}=7$ TeV collision energy in $pp$-frame.  Before the scheduled shut-down at  the end of 2012, the accumulated  luminosity is expected to be  more than $10 {\rm fb}^{-1}$. At full power, the LHC is designed to produce luminosity greater than $10-100 {\rm fb}^{-1}$ per year with a center of collision energy  of 14 TeV.

For the black hole, the basic observable quantity is the  number of  specific ``black hole signals'', which can be expressed as:
\begin{eqnarray}
N_{\text{BH signal}} = {\cal L}_{\text{\tiny LHC}} \times \sigma (pp \to \BH +X) \times Br(\BH \to \text{signal}),
\end{eqnarray}
where ${\cal L}_{\text{LHC}}$ is the luminosity of $pp$ collision, which is accumulated by the LHC at a given time, $\sigma(pp\to \BH+X)$ is the production cross-section for BH production  and $Br(\BH \to \text{signal})$ is the branching ratio of the decaying black hole to the signal.

The production cross-section of BH depends on the collision energy as previously discussed.  The greater the collision energy is, the larger the cross-section: $\sigma(E) \sim (G_D E)^{2/(D-3)}$ for semi-classical black holes. Thus, a higher energy run with a larger collision energy  ($\sqrt{s}=14$  TeV) will have an obviously better chance of producing black holes than the current 7 TeV run. If possible, a future high energy collider with even higher CM energy, e.g., the VLHC  with 100 TeV CM,  is definitely more desirable. Giddings and Thomas (2001) \cite{Giddings:2001} and Dimopouplos and Landsberg (2001) \cite{Dimopoulos:2001} estimated $N_{\rm BH}$ and concluded that the black hole production rate could be large, with $\md\approx1$ TeV, but later developments also showed that there could be significant suppression of the rate. The most important factor is the threshold of black hole production.  For a thermal or classical black hole, the energy threshold should be much larger than $\md$ or $\sqrt{s}\gg \md$, otherwise quantum gravitational effects easily spoil the theoretical predictions. Based on the entropy counting, for instance, the threshold should be set to at least $M_{\rm min}/\md > 5$ for semi-classical black holes. With this threshold, the cross-section is largely suppressed by PDF as shown by Cheung \cite{Cheung}.

After a black hole is formed, it decays  and emits Hawking radiation.  The actual decay products depend on the number of extra dimensions, the mass, charge and angular momentum of the black hole, the wave function profile (the location) of particles in extra dimensions and also the geometry of the extra dimension. However, the large entropy of a black hole and the basic property of Hawking radiation determine some robust properties of the signature:
\begin{itemize}
  \item Large multiplicity
  \item Flavor blindness
\end{itemize}

The large multiplicity of the signal is inherited from the large entropy of the thermal black hole. Thermal black holes typically have a large mass ($\gg \md$) and the Hawking radiation contains large numbers of particles, jets, in particular. The flavor blindness of the black hole signal can be understood since the Hawking radiation is essentially thermal.  Even when we take the greybody factors into account (which we will discuss later), flavor blindness remains. Statistically we would expect the same number of electron, muon and tau particles in the Hawking radiation. However, Hawking radiation depends on the spin of the particle and also on the number of degrees of freedom for each particle. For instance, a gluon has eight colorful degrees and the spin is one. An electron, on the other hand, does not carry any color charge and its spin is $1/2$ such that the portion in Hawking radiation is relatively smaller than a gluon.

\subsection{The production cross section in a $pp$ collision}

\begin{figure}[t]
\begin{center}
        \includegraphics[width=.75\textwidth]{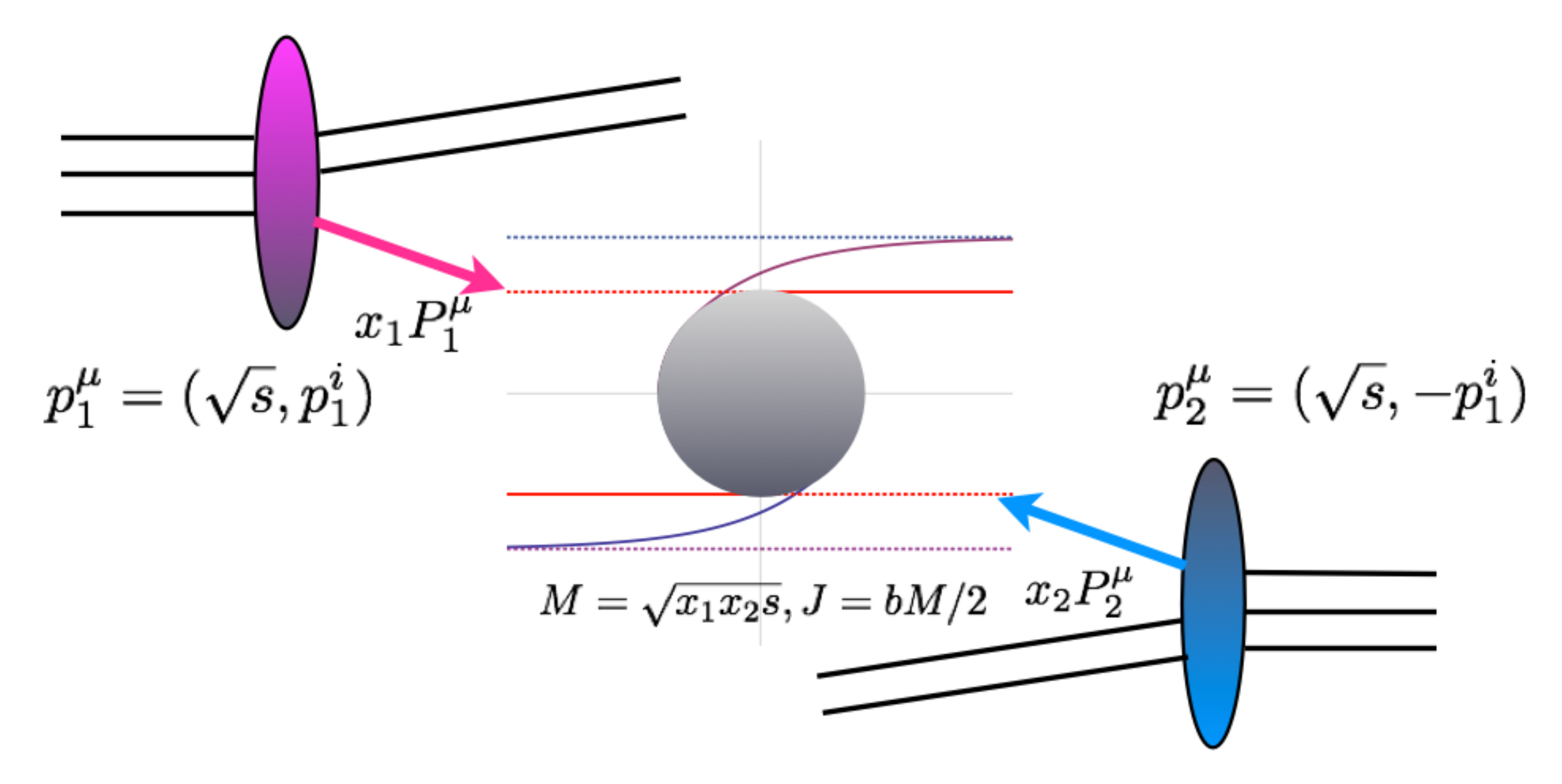}
        \end{center}
  \caption{A schematic view of black hole formation in $pp$ collision. }
\label{fig:parton}
\end{figure}

The LHC is a proton-proton ($pp$) collision machine.  Each proton is composed of several  partons, and  each parton carries a part of the proton energy and interacts with other partons or passes by as a spectator at the moment of collision. The actual collision energy therefore is less than the total energy of protons, but a part of it.  Fig. (\ref{fig:parton}) is  a schematic of $pp$ collision that produces a black hole of mass $\sqrt{x_1x_2 s}$ and angular momentum $J$. Considering the parton distribution function (PDF),  and the $f_i(x,Q^2)$ of each parton ($i$) \cite{MSTW, CTEQ}, the hadron level cross-section is obtained as:
\begin{eqnarray}
\sigma_{pp\to {\rm BH} + X} =\int_{\md^2/s}^1 du \int_u^1 \frac{dv}{v} \sum_{ij} f_i(v, \hat{s}) f_j(\frac{u}{v},\hat{s}) \hat{\sigma}_{ij \to {\rm BH}},
\end{eqnarray}
where $\hat{\sigma}_{ij \to {\rm BH}}$ is the cross-section of the two partons ($i$ and $j$). The sum runs $i,j =\left(q, \bar{q}, g\right)$ and $q=(u,d,s,c,b)$, i.e., partons in a proton.

\subsubsection{Form factor}

The parton level cross-section of black hole production is formally given as a geometric cross-section modified by a ``form factor'' ${\cal F}$:
\begin{eqnarray}
\hat{\sigma}_{a b \to {\rm BH}} = {\cal F} \pi r_s(\sqrt{\hat{s}})^2,
\end{eqnarray}
where $r_s$ is the Schwarzschild radius of the given CM energy $\sqrt{\hat{s}}$ (see Eqs. \eqref{eq:rs} and \eqref{eq:kn} for explicit expressions).  Here the hatted variables are parton level quantities. In general,  $\sqrt{\hat{s}}$, the collision energy  provides the upper limit of the produced black hole: $M \leq \sqrt{\hat{s}}$ considering the possible energy loss before and during the formation process. The form factor ${\cal F}$ implicitly includes information about energy loss, angular momentum, charge and other dynamical modifications.

\begin{figure}[h]
\begin{center}
    \includegraphics[width=.55\textwidth]{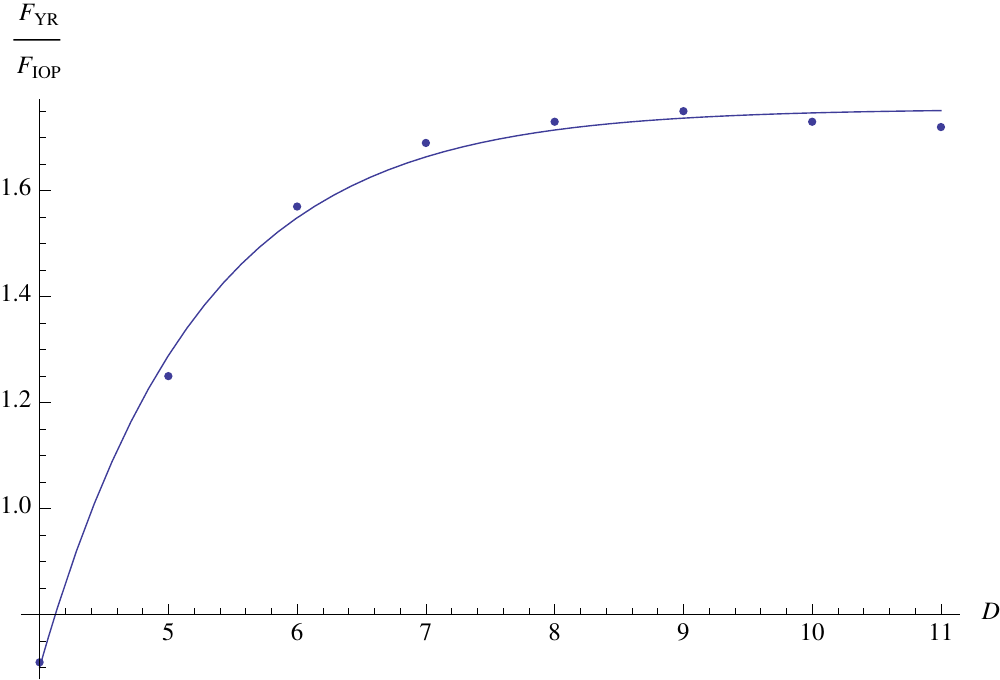}
  \caption{The fitted function for YR and IOP. The dotted points are numerical values in YR and the
  solid curve is the fitting curve $\frac{{\cal F}_{\rm YR}}{{\cal F}_{\rm IOP}}=(1.75451 - 27.7194 e^{-0.817237 D})$ in $4\leq D \leq11$.}
\label{fig:fit}
\end{center}
\end{figure}

A commonly used expression for the form factor is obtained following the relation given in Eq. \eqref{eq:bmax}:%
\begin{eqnarray}
{\cal F}_{\rm IOP}\equiv \left[\frac{2}{1+\left(\frac{D-2}{2}\right)^2}\right]^{\tfrac{2}{D-3}}.
\end{eqnarray}
This form factor describes the geometric cross-section of a `black disk' taking the angular momentum of the black hole into account \cite{Park:2001, IOP1}.

Another commonly used form factor is  from a numerical work  by Yoshino-Rychkov (2005) \cite{formation3} using the Aichelburg-Sexl solutions discussed in the previous section.  Here we provide a useful approximate formula to fit ${\cal F}_{\rm YR}$ to ${\cal F}_{\rm IOP}$  (see Fig. \ref{fig:fit}):
\begin{eqnarray}
{\cal F}_{\rm YR} &=& \left(a_1 - a_2 e^{-a_3 D}\right){\cal F}_{\rm IOP}, \\
(a_1,a_2,a_3)_{\rm fit}&=&(1.75451, 27.7194, 0.817237) \nonumber.
\end{eqnarray}

Assuming $\sqrt{\hat{s}}=10$ TeV, $\md=1$ TeV and ${\cal F}={\cal F}_{\rm IOP}$ as reference inputs, the parton level cross-section  for $D=5-10$ dimensions is roughly $\hat{\sigma} \approx (3.2, 6.4, 10.3, 14.7, 19.7, 25.1)$ nb.

\subsubsection{Split fermion, spin, charge and other factors}

\begin{figure}[h]
\begin{center}
    \includegraphics[width=.45\textwidth]{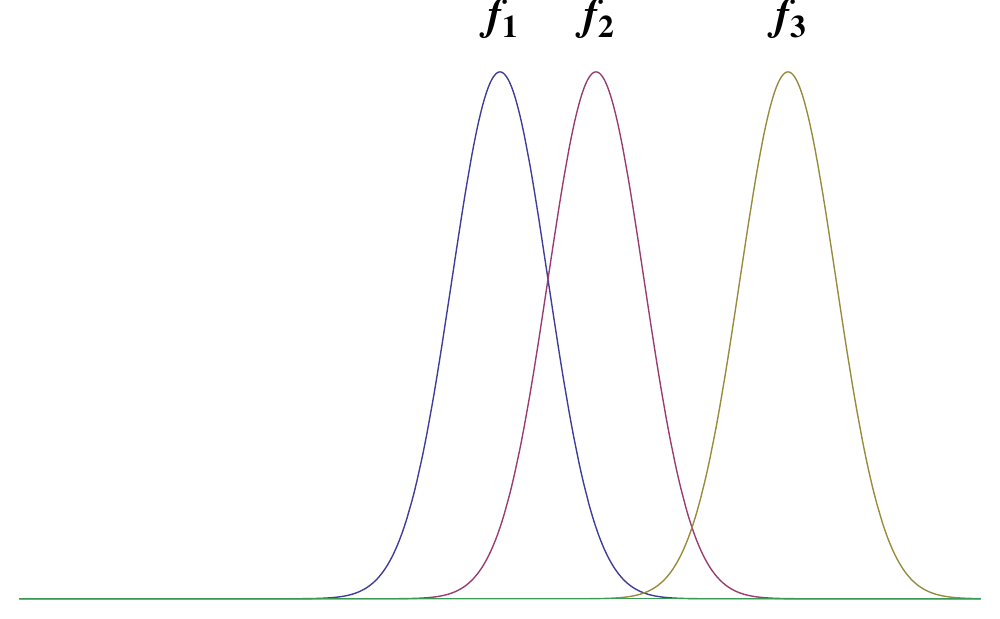}
  \caption{Split fermions in extra dimensions. $f_1$ and $f_2$ are rather closed located but $f_3$ 
  is at a distance. With this splitting, the probability of making a black hole is weighted by the wave function overlap integral.}
\label{fig:split}
\end{center}
\end{figure}

One should note that the parton level cross-section can be largely modified if fermions are {\it split} in extra dimensions \cite{ArkaniHamed:1999}.  The splitting is a unique phenomenon in extra dimension models and it may be responsible for the observed Yukawa hierarchy \cite{Mirabelli:1999} (also see \cite{Park:2009, Csaki:2010} for universal extra dimension (UED) setup) and the proton longevity problem (Adams (2000) \cite{Adams:2000})  in particle physics. 
In Fig. \ref{fig:split}, we drew three fermions in an extra dimension with splitting. The wave function of $f_1$ has a relatively larger overlap with $f_2$ than with $f_3$. The location of the wave function is controlled by the bulk mass. The wave functions  $f_1$ and $f_2$ are relatively closer, but $f_3$ is a bit more separated. As a result, the wave function overlap between $f_1$ and $f_2$ is larger than that between $f_{1,2}$ and $f_3$.

To take the splitting into account, Dai et al (2006) \cite{Dai:2006} introduced an extra dimensional weighting parameter $w^{ij}(b)$ for a given impact parameter $b$ on `our brane':
\begin{eqnarray}
w^{ij}(b)\equiv \int_0^{\sqrt{b_{\rm max}^2-b^2}} P^{ij}(b_e) d b_e
\end{eqnarray}
where $P^{ij}(b_e)$ is the probability that the particles' separation in the extra dimensions will be $b_e$.
With the weighting parameter, the total cross-section can be expressed as
\begin{eqnarray}
\sigma_{pp \to {\rm BH}+X} =\sum_{ij}\int_{\md^2/s}^1 du \int_u^1 \frac{dv}{v} \int_0^{b_{\rm Max}} db 
w^{ij} 2\pi b  f_i(v,Q)f_j(u/v,Q),
\end{eqnarray}
which can be significantly suppressed by small overlaps among  split fermions. The splitting can also be  important in warped extra dimension (see e.g. Rizzo (2006) \cite{Rizzo:2006}).


Hoop conjecture does not indicate whether  the spin, charge and other quantum numbers of a collapsing  or colliding particle may cause differences in black hole formation, but we at least know that on  some dynamical time scale they may cause differences. Yoshino and Mann (2006) \cite{Yoshino:2006dp} and Gingrich (2007) \cite{Gingrich:2006} studied the effect of charged partons on black hole production and  concluded that the black hole cross-section can be significantly reduced (also see Yoshino et.al. (2007) \cite{Yoshino:2007a}).  In most papers on black holes at the LHC, it is assumed that the extra dimension is asymptotically flat and large so that the Schwarzschild-Tangherlini or Myers-Perry black hole provide a reasonably good description of the black hole.
However, `brane' may carry non-vanishing tension and the extra dimension may have a finite curvature in the vicinity of the produced black hole. These effects can also affect the production rate, but quantitative details are not  known.


\subsection{Black hole decay by Hawking radiation}

\begin{figure}[h]
\begin{center}
    \includegraphics[width=.55\textwidth]{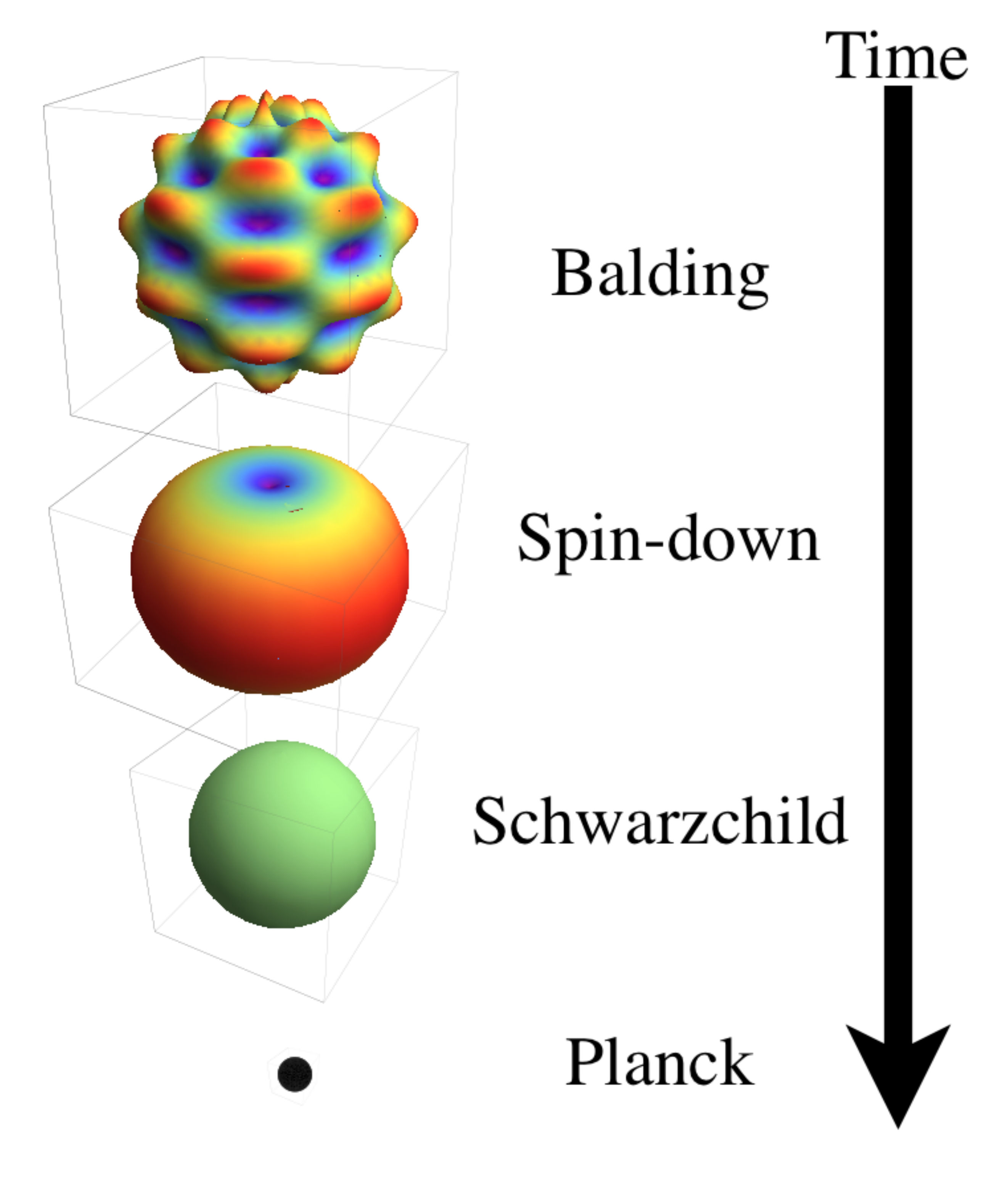}
  \caption{A black hole evolves via four steps-balding, spin-down, and the  Schwarzschild and Planck phases. The balding phase is a dynamical stage of black hole formation during which gravitational radiation  
 takes the energy and angular momentum away, but the precise rates are still  unknown. The spin-down phase and Schwarzschild phase are essentially described by Hawking radiation, then the final Planck phase meets
 the quantum gravitational uncertainty.}
  \end{center}
\label{fig:time}
\end{figure}

In the first stage of production, the concentrated energy is not settled down to a simple, symmetric black hole,  but  rather violently emitting gravitational waves. This stage is called the {\it balding phase} after the `no-hair' theorem. The {\it spin-down phase} follows after the balding phase. In the spin-down phase, the black hole is well described by the MP solution with one rotation parameter. In this stage, Hawking radiation takes energy and angular momentum away so that  the rotation of the black hole  slows down and proceeds to the non-rotating phase or the {\it Schwarzschild phase}. In the Schwarzschild phase, the rest of the energy is emitted as Hawking radiation, which is rather symmetric and democratic (equal for each species). At the end of the Schwarzschild phase, the mass of the black hole becomes not much greater than $\md$ for which a full quantum gravitational description is required.  We schematically illustrate the evolution of the black hole produced in particle collision  in Fig. \ref{fig:time} including the balding, spin-down, Schwarzschild and the final {\it Planck phase} where $M\sim \md$.  The phases were first described in \cite{Giddings:2001}.

\subsubsection{Black hole evolution: Balding, spin-down and the Schwarzschild phase}

A black hole, that is first formed in a particle collision is expected to be highly asymmetrical. Considering the   violent environment of  colliding and merging particles, no  particular symmetry  is expected for a black hole  in the early stage of production. Thus,  in the  balding phase, the black hole loses its ``hair" by emitting energy, angular momentum and charge in the form of gravitational radiation and gauge field emission.  At the end of this phase,  it will settle down to  a stationary, rotating black hole then it can be nicely described by the Myers-Perry solution. An upper bound for the angular momentum was estimated by taking the initial condition of the angular momentum into account using the hoop conjecture \cite{Park:2001}: $0 < a\equiv (n+2)J/(2 M_{bh}R_{bh})<(n+2)/2$, but there is a much room for studies aiming toward understanding the balding phase. One may hope that numerical studies for dynamical black hole formation can achieve an improved understanding in the future (see Sec. \ref{sec:numerical}).

\begin{figure}[h]
\begin{center}
    \includegraphics[width=.45\textheight]{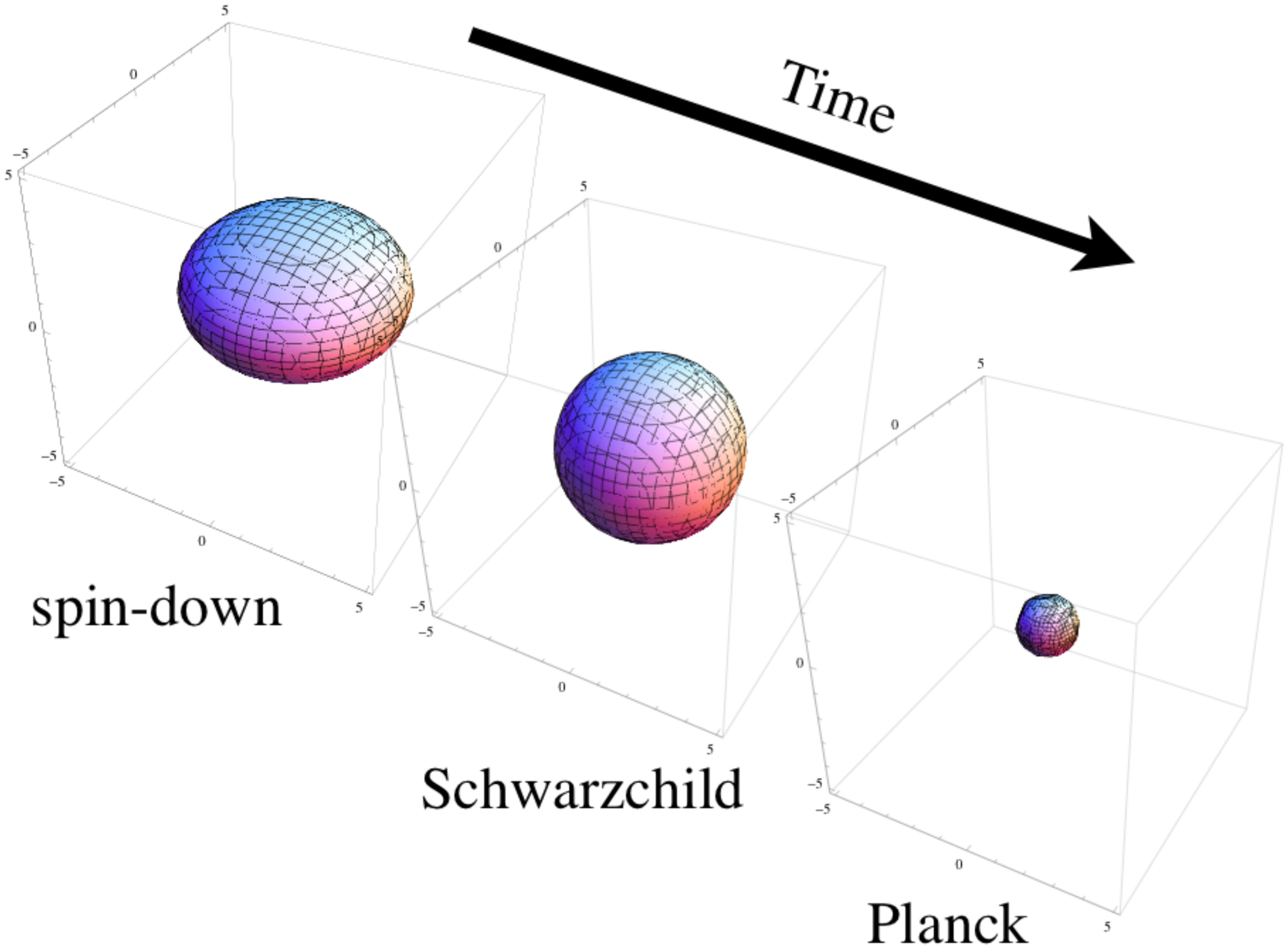}
  \caption{Evolution of black hole via Hawking radiation. The geometry becomes more spherically symmetric
  in spin-down phase. When the angular momentum is lost completely, Schwarzschild phase takes over then black hole
  evolves toward the Planck phase.}
  \label{fig:evolution}
 \end{center}
\end{figure}

The phase following the balding phase is the spin-down phase. During the spin-down phase, the black hole becomes more symmetric as it loses its angular momentum. As seen in  Fig. \ref{fig:evolution}, the apparent horizon geometry becomes more spherically symmetric by emitting angular momentum and proceeds to the Schwarzschild phase.

 In spin-down and the Schwarzschild phases, the black hole loses its energy and angular momentum (when $J>0$) through Hawking radiation ~\cite{Hawking:1974}:
 \begin{eqnarray}
\frac{d^2 }{dt d\omega} \binom{E}{J}=\frac{1}{2\pi} \sum_{jm} \frac{ \Gamma_{s,\ell, m}(\tilde{\omega}, a_*) }{e^{\tilde{\omega}/T_h} -(-1)^{2s} } \binom{\omega}{m}, 
\end{eqnarray}
where
\begin{eqnarray}
\tilde{\omega} = \omega - m \Omega, \\
\Omega = \frac{a_*}{(1+a_*^2) r_h},\\
\end{eqnarray}
are the reduced energy and the angular velocity on the horizon, respectively. The numerical factor ${}_s\Gamma_{\ell,m}$ is called the {\it greybody factor} (also known as the transmission coefficient), which describes the modification of the radiation in the vicinity of the horizon due to the curved geometry. Indeed, taking the greybody factor into account, the spectrum of Hawking radiation to be measured by an observer at infinity does not look quite black-body like.  The greybody factor for a given spin-$s$ of the radiating particle  is defined to be the rate of absorption  by the hole for a steady in-falling flux at  infinity,
\begin{eqnarray}
{}_s\Gamma_{\ell, m}\equiv \frac{{}_s\dot{N}_{\ell m}^{\rm in}-{}_s\dot{N}_{\ell m}^{\rm out}}{{}_s\dot{N}_{\ell m}^{\rm in}},
\end{eqnarray}
under the boundary condition that the radial wave is purely ingoing at the horizon. 
The greybody factor  can be found by solving the generalized Teukolsky equation in Eq. \eqref{eq:radial}  and the spin-weighted spheroidal harmonics in Eq. \eqref{eq:angular}.

The generalized Teukolsky equation in $D\geq 4$ was first derived by Ida-Oda-Park (2002) \cite{IOP1}  utilizing the Newman-Penrose formalism developed in \cite{Newman:1961, Penrose:1985}. The equation turned out to be separable as the projected MP metric is Petrov type D.  For a $D=4$ Kerr black hole, the original Teukolsky equation was derived and solved in the early 1970s  by Teukolsky \cite{Teukolsky:1973a, Teukolsky:1973b, Teukolsky:1974} and Page \cite{Page:1976a, Page:1976b}. A recent work demonstrated that exact solutions for the $D=4$ Teukolsky equation can be found (see, e.g.,  Fiziev (2010) \cite{Fiziev:2009}).  The radial equation of the generalized Teukolsky equation for $D=4+n$ dimension \cite{IOP1} is
\begin{eqnarray}
&&\Delta^{-s} \frac{d}{dr} \left(\Delta^{s+1} \frac{dR}{dr}\right) \label{eq:radial}\\
&&+\left[\frac{K^2}{\Delta}+s\left(4i\omega r -i \frac{K\left(2r + (n-1)\mu r^{-n}\right)}{\Delta} -n(n-1)\mu r^{-n-1}\right)+2 m a \omega - a^2 \omega^2 -A \right]R =0, \nonumber 
\end{eqnarray}
where $\Delta(r)= r^2 +a^2 -\mu r^{-n+1}$, $K=(r^2+a^2)\omega -ma$ and $A$ is the angular eigenvalue, which is an  integration constant coming from the angular equation. The angular equation is called the spin-weighted spheroidal harmonic equation, which is common for $D=4$ and $D=4+n$ (see Goldberg et al. (1966)\cite{Goldberg:1966},  Leahy, Unruh(1979) \cite{Leahy:1979}, Leaver(1985)\cite{Leaver:1985}, Seidel (1988) \cite{Seidel:1989}
and Casals et al. (2009) \cite{Casals:2009}):
\begin{eqnarray}
\frac{1}{\sin \theta}\frac{d}{d\theta}\left(\sin\theta \frac{d S}{d\theta}\right)+\left[(s- a \omega \cos\theta)^2 -(s \cot \theta + m \csc \theta)^2 - s(s-1)+ A \right]S=0.
\label{eq:angular}
\end{eqnarray}
%

\begin{figure}
\begin{center}
    \includegraphics[width=.69\textwidth]{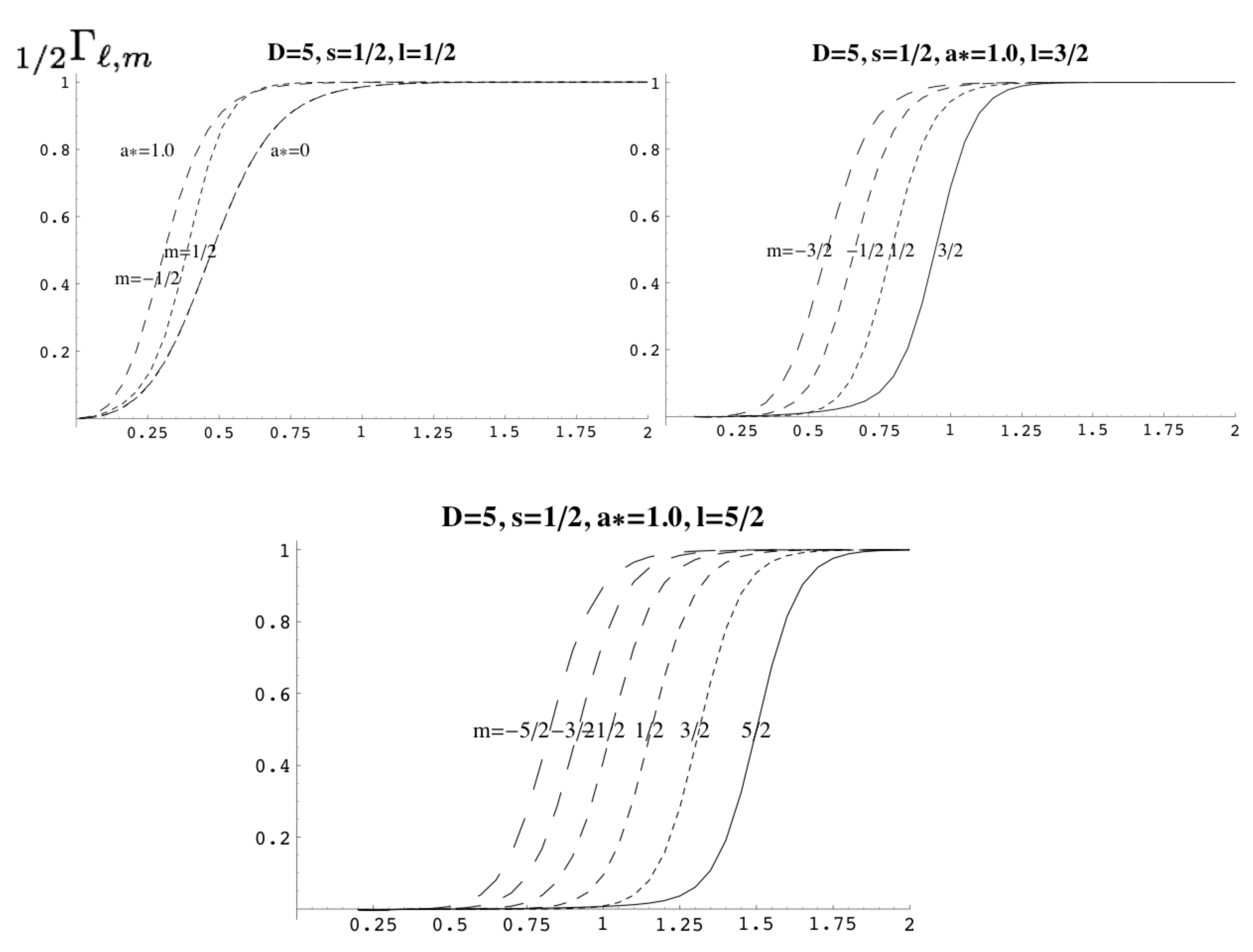}
  \caption{ Greybody factors for a fermion field in $D=5$.}
  \label{fig:gray5D}
  \end{center}
\end{figure}

\begin{figure}
\begin{center}
    \includegraphics[width=.69\textwidth]{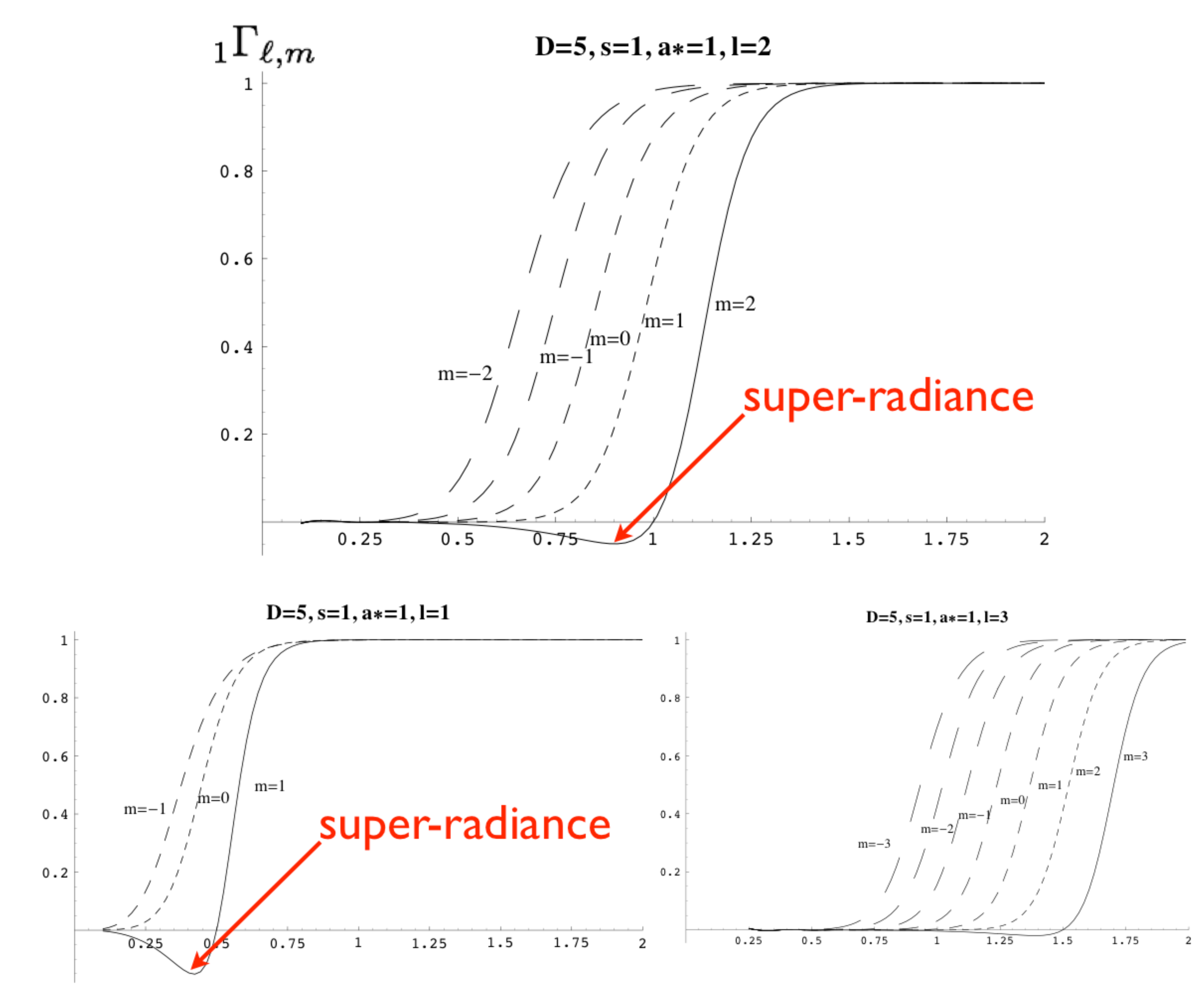}
  \caption{Greybody factors for $D=5$, vector field. Super-radiance modes (i.e., $\Gamma<0$)
  are indicated. }
  \label{fig:gray5Dv}
  \end{center}
\end{figure}

\begin{figure}
\begin{center}
    \includegraphics[width=.89\textwidth]{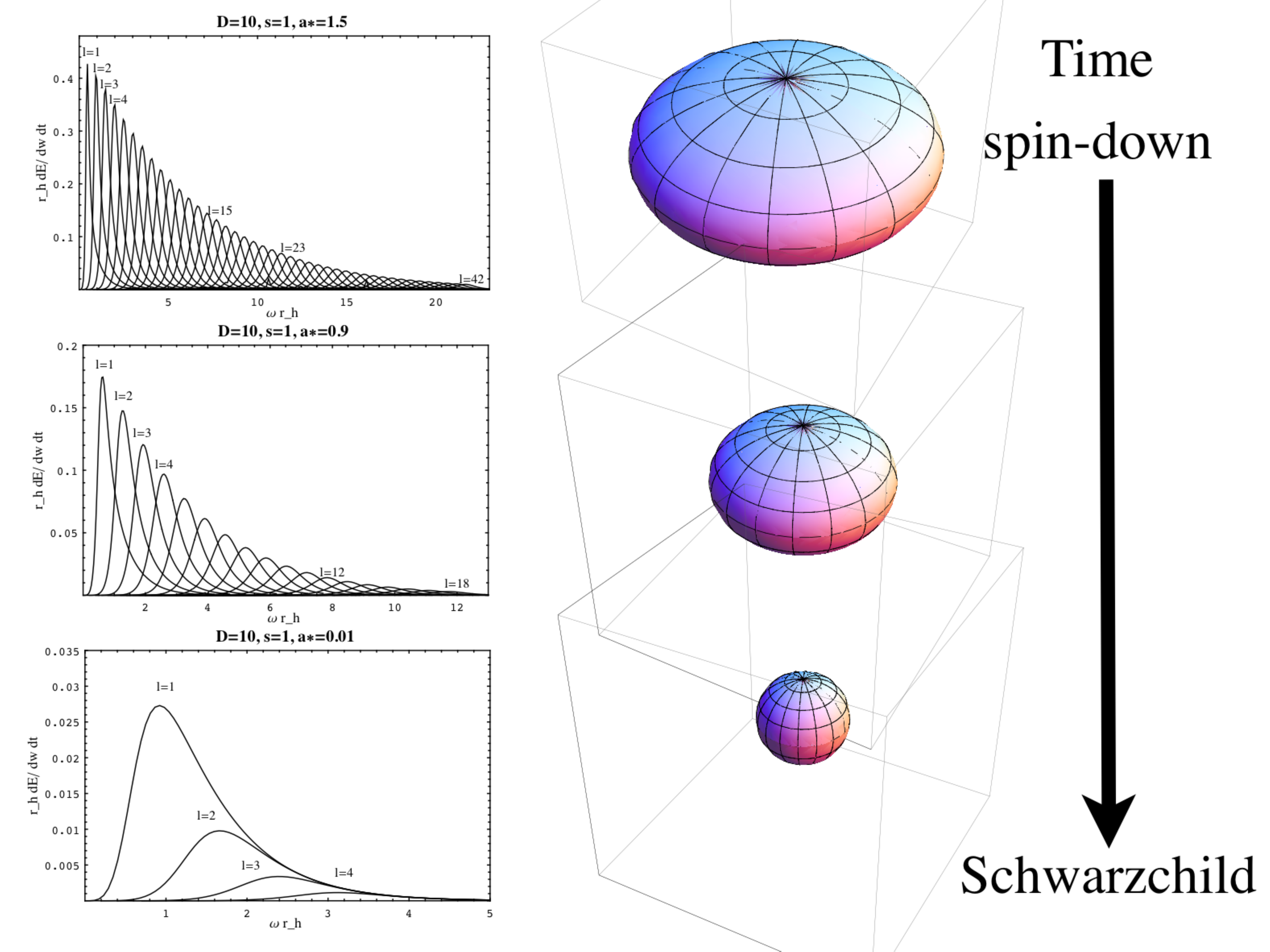}
  \caption{ The power spectrum of the $D=10$  dimensional black hole to the brane
  vector field and a schematic of horizon geometry. The highly rotating black hole emits Hawking radiation 
  and settles down to a non-rotating black hole. When the black hole is non-rotating, the spectra are `simple' and `calm' compared to the violent and exciting radiation from a highly rotating black hole with many angular modes with several $(l,m)$. By reconstruction of the spectrum, in principle, one can deduce
  the angular momentum and mass of the decaying black hole.}
  \label{fig:10Ds1}
  \end{center}
\end{figure}

At low energy expansion, in the regime $\omega r_h \ll 1$, the analytic form of the greybody factors for fields with spin-$s=0,1/2, 1$ were found  by Ida-Oda-Park (2002) \cite{IOP1} using the `matching method': the radial equation is solved near the horizon and at infinity then matched in the overlapping region. The matching condition determines the wave function in the whole regime in $r/r_h \in (1, \infty)$. (For a non-rotating black hole in $D>4$ at the low energy limit of the radiation see \cite{Kanti:2002NR, Harris:2003}).
However at high energy, $\omega r_h \gsim 1$ , the matching method does not work so that  one has to rely on numerical methods. The first numerical solution for a scalar ($s=0$) for the whole energy regime, $ \omega r_h \in (0, \infty)$, was found by Ida-Oda-Park (2004) \cite{IOP2-1, IOP2-2} and  later independently by Harris, Kanti (2005) \cite{Harris:2005} and  Duffy et al. (2005) \cite{Duffy:2005}. In \cite{IOP2-1,IOP2-2}, the super-radiance modes in $\tilde{\omega}=\omega-m \Omega <0$ are explicitly found for the first time for the MP black holes (see Fig. \ref{fig:10Ds1}).  In the winter 2005-2006, Casals et al. (2005) \cite{Casals:2005} found  the first numerical results for vector field ($s=1$) and  Ida-Oda-Park (2006) \cite{IOP3} found numerical results for a vector as well as a spinor ($s=1/2, 1$). Casals et al. (2006) \cite{Casals:2006} also got similar results for $s=1/2$.  Even though the effect of the mass and charge of radiated particles is expected to be {\it small} for semi-classical black holes, $M_{\rm bh}\gg \md$, for which we can only neglect the back reactions to the black hole geometry, some related studies have been conducted  for massive (charged) scalar emission by Casals et al. (2008) \cite{Casals:2008s}, Sampaio (2009, 2009)  \cite{Sampaio:2009tp, Sampaio:2009ra} and Kanti et al. (2010) \cite{Kanti:2010mk}, for bulk massive scalar by a tense brane by Kobayashi et al. (2007) \cite{Kobayashi:2007zi} and for massive fermion by Rogatko et al. (2009) \cite{Rogatko:2009jp}. Also a part of gravitational radiation (to tensor mode) was studied by Cornell et al. (2006) \cite{graviton1}, Cardoso et al. (2006) \cite{graviton2} and Kanti et al. (2009) \cite{graviton3}.

 In Ida-Oda-Park (2006) \cite{IOP3}, the results for the time evolution of the MP black hole were found and the amount of energy loss through the Hawking radiation to the complete set of the standard model particles (quarks, leptons, Higgs boson and gauge bosons) was calculated for the spin-down and Schwarzschild phases. The result is expected to be valid for a lower dimensional case ($D=5$), especially if the Kaluza-Klein gravitons are heavy as in the Randall-Sundrum model,  
but a sizable modification is expected in higher dimensional cases ($D\geq 6$) due to a sizable gravitational radiation. Below is a summary of the time evolution of a rotating black hole (see Fig. \ref{fig:evolution} for a schematic of black hole evolution).
\begin{itemize}
\item Generically, the faster the black hole rotates, the larger the rate of the angular momentum radiation. Therefore, fast angular momentum loss happens during the early stage of evolution, then the phase with a rather low emission rate persists until the angular momentum is completely radiated away and the black hole stops rotating. 
During this {\it spin-down phase}, typically more than 70\% or 80\% of a black hole's mass is lost before the angular momentum parameter becomes very small ($< 1\%$ of the initial value) depending on the number of extra dimensions.

\item The most effective loss of the angular momentum is caused by vector field radiation.  Compared to the spinor and scalar field, the vector field emission dominates the radiation when a black hole rotates fast. Thus, the Hawking radiation is not equal for each spin state. In the standard model, the gluon (with eight colors) and weak gauge bosons ($W^\pm, Z^0$) and photon ($\gamma$) contribute and dominate over fermions (quarks and leptons) and the Higgs boson.
\item   For a typical event with a large initial angular momentum, more than half of a black hole mass is emitted when the hole is still rotating such that the Hawking radiation would be anisotropic rather than isotropic. 
\item At the end of the decay, the mass of the black hole will reach the Planck scale: $M \sim \md$. This domain is not controlled by semi-classical GR. A sizable back-reaction due to the discontinuous radiation of particles could also be important in this domain as the mass of the black hole is not large. 
\end{itemize}

A theoretical issue of separability of the Hamilton-Jacobi equation and especially the Klein-Gordon equation is important to predict the detailed behavior of fields in the black hole background. The MP black hole posseses spacetime symmetries generated by the Killing vectors and  also admits the antisymmetric Killing-Yano and symmetric Killing tensors. The spacetime symmetry of the $D = 2n + 1 + \epsilon$ dimensional MP metric guarantees the existence of $n+2$ integrals of motion including the normalization of the velocity.  To ensure separability,  at least $n=D-(n+2)$ additional integrals of motion must exist. For $D=4$ and $D=5$ one has $n=1$, thus the Killing tensor that exists in the MP metrics is sufficient for the separation of variables. For $D \geq 6$, $n> 1$, such that one cannot expect the separation of variables unless additional hidden symmetries exist \cite{Frolov:2006hd, Frolov:2008hd}.

\subsubsection{Small ADD black hole and RS black hole}

\begin{figure}[h]
\begin{center}
    \includegraphics[width=.65\textwidth]{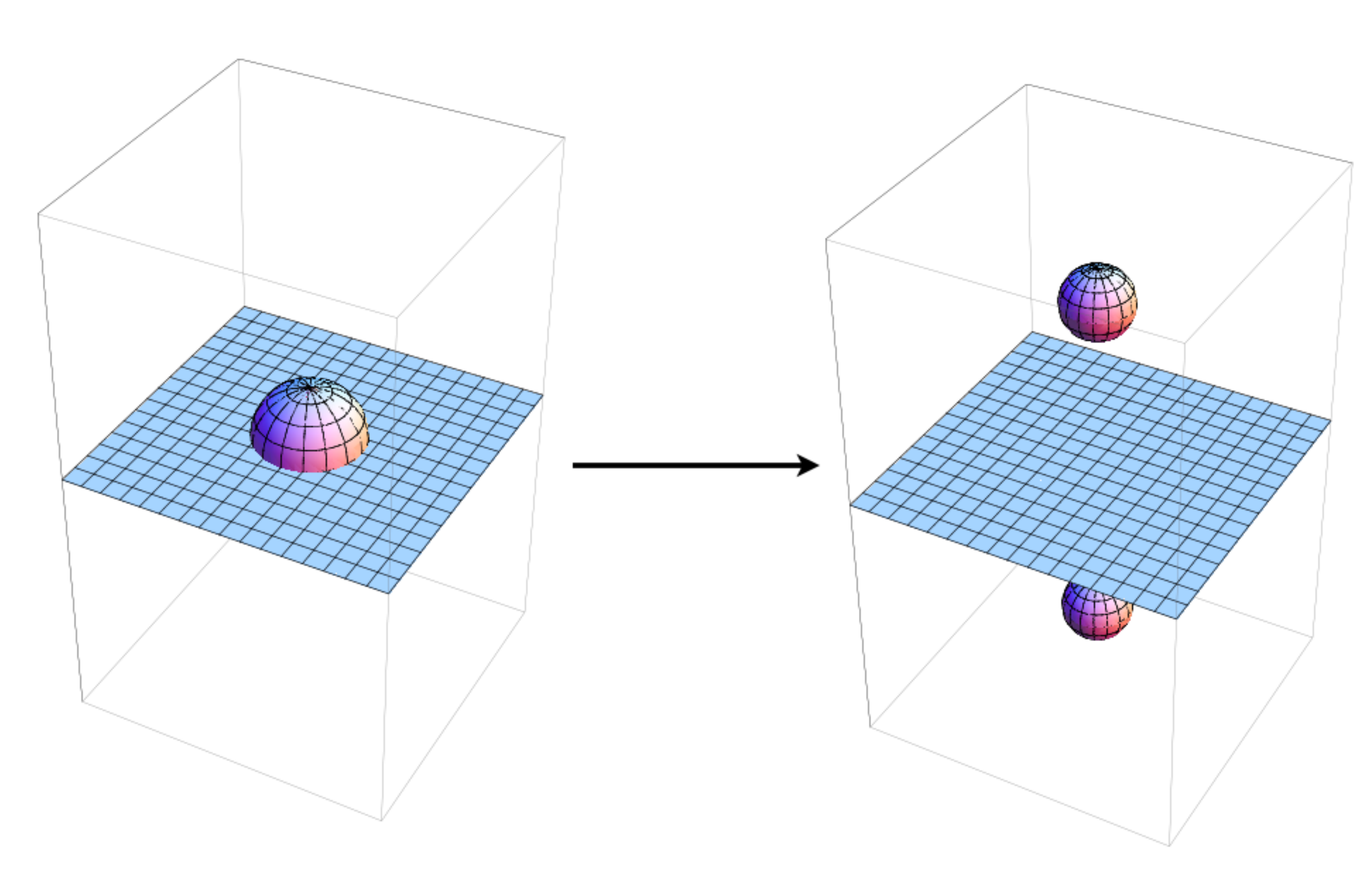}
  \caption{A black hole on the brane recoils with  bulk radiation. This process is forbidden in the presence of $\mathbb{Z}_2$ symmetry about the brane position. A black hole of mass $M$ should be split into two smaller black holes with mass $M/2$ for each if $\mathbb{Z}_2$ is exact. However, the entropy of the initial black hole  is $S_{\rm ini}=S(M)\propto M^{(2+n)/(1+n)}$ and the entropy for the final black holes is $S_{\rm final}=2 S (M/2) \propto 2 (M/2)^{(2+n)/(1+n)} < S_{\rm ini}$,  {\it smaller} than the original entropy. Thus, such a process is forbidden by the area theorem.}  
  \end{center}
\label{fig:brane_bh}
\end{figure}

If a black hole is small enough (i.e., $r_h \ll r_c$) the small black hole can be regarded as a higher dimensional black hole with $S^{2+n}$ event horizon topology in $4+n$ dimensions. As long as the curvature radius is large compered to the size of the event horizon, there seems to be no qualitative distinction between a black hole in the ADD background and the one in the RS background; both are well approximated by MP black solutions. However, it should be noted that RS geometry has a $\mathbb{Z}_2$ reflection symmetry about the `brane' where the mini black hole is likely to be produced such that the behavior of a black hole in different scenarios might be distinguishable \cite{Stojkovic:2004}. A small ADD black hole (i.e., a black hole in the ADD background) radiates mostly to the bulk in the first phase of production, particularly due to the large angular momentum. A black hole can recoil and leave the brane \cite{Frolov:2002, Frolov:2002b, Frolov:2004a, Frolov:2004b}. However, for a small RS black hole (i.e., a black hole in the RS background), bulk radiation is strongly suppressed due to the small number of bulk degrees of freedom and the presence of $\mathbb{Z}_2$ symmetry. A black hole cannot recoil and leave the brane. 

One can easily understand such a phenomenon. Consider a black hole on a brane as in Fig. \ref{fig:brane_bh}. In the presence of $\mathbb{Z}_2$ in RS, recoil of the black hole is identified with splitting of the black hole into two smaller black holes in opposite directions. Each smaller black hole has a half the mass of the original black hole. Now considering the entropy of the initial and final states, one can conclude that the final state has a smaller entropy than the original entropy, which is forbidden by the area theorem (the black hole area never decreases). As the recoil and bulk radiation of graviton should be regarded as non-visible or missing energy signatures at particle detectors, ADD and RS black holes can be distinguished by the amount of missing energy signatures, for instance. However the details of the signature could quite possibly depend on the details of brane configuration and brane tension, which should be clarified by further studies. The role of more general warped compactifications also remains to be understood.


\subsection{Recent CMS results}

Searches for microscopic black hole production and decay in $pp$ collisions at a CM energy $7$ TeV have been conducted and reported by the LHC (CMS collaboration) using $35~ {\rm pb}^{-1}$ \cite{CMS} and $1~ {\rm fb}^{-1}$ \cite{CMS2} (CMS physics analysis summary (CMS PAS EXO-11-071) ) data sets  obtained before the summer of 2011. More data has been collected by the CMS as well as ATLAS so that new results are expected soon. The higher energy upgrade from $7$ TeV to $14$ TeV of the LHC is expected, and we will definitely will be able to learn more about the physics on TeV scales  in the coming years. The higher energy upgrade is particularly important for the discovery of the new physics models with a high energy threshold. 

For the black hole search at the LHC, the following set of criteria and assumptions on signal and background  has been considered in the CMS experiment \cite{CMS, CMS2}:
\begin{itemize}
\item The ADD model is assumed for low scale gravity with $\md \sim 1$ TeV. The produced black hole is confined on a brane where {\it all} particles except gravitons reside.
\item A semi-classical black hole has been considered even though a large quantum correction is expected (we will discuss this more in the later part of this section).
\item High multiplicity ($N\geq 2-8$) from the black hole with a large entropy and  democratic, highly isotropic Hawking radiation, which is typical of a non-rotating black hole,  is taken into account for signals. 
\item  For a trigger, the final state particles (mostly jets and hadrons) carrying more than $100$ GeV of energy are considered. The large scalar sum of the transverse energy, $H_T =\sum_{jet} E_T \geq 50(100-200)$ GeV, with jet $E_T\geq 10 (20-30) $ GeV at the first level trigger (High level trigger) is used.  For event selection, a large scalar sum of the $E_T$ on the $N$ individual objects including jets, electrons, photons and muons, ($\equiv \sum_{E_T>50 {\rm GeV}} S_T \geq (1.5-4.0)$ TeV), are considered for various $M_D \in (1.5, 3.0)$ TeV in $D\in[6,10]$.  (note: $M_D =\md \times (8\pi/{\cal N}_n)^{1/(n+2)}$ and ${\cal N}_n =(2\pi)^n$ in the PDG unit.)
\item  Black hole signal events are simulated using the parton level BlackMax generator \cite{Blackmax} (version 2.01.03), followed by a parton showering fragmentation with \texttt{PYTHIA}  \cite{Pythia} (version 6.420) and a well validated simulation of the CMS detector response.
\item  The main background to black hole signals  are from QCD multi-jet events, $W/Z+{\rm jets}$ and $t\bar{t}$ production were estimated from Monte-Carlo simulations using \texttt{MADGRAPH} \cite{Madgraph} with the CTEQ6L PDF set \cite{CTEQ} followed by \texttt{PYTHIA} parton showering and full CMS detector simulation via \texttt{GEANT4} 
\cite{Geant4}.
\end{itemize}

\begin{figure}[h]
\begin{center}
    \includegraphics[width=.32\textwidth]{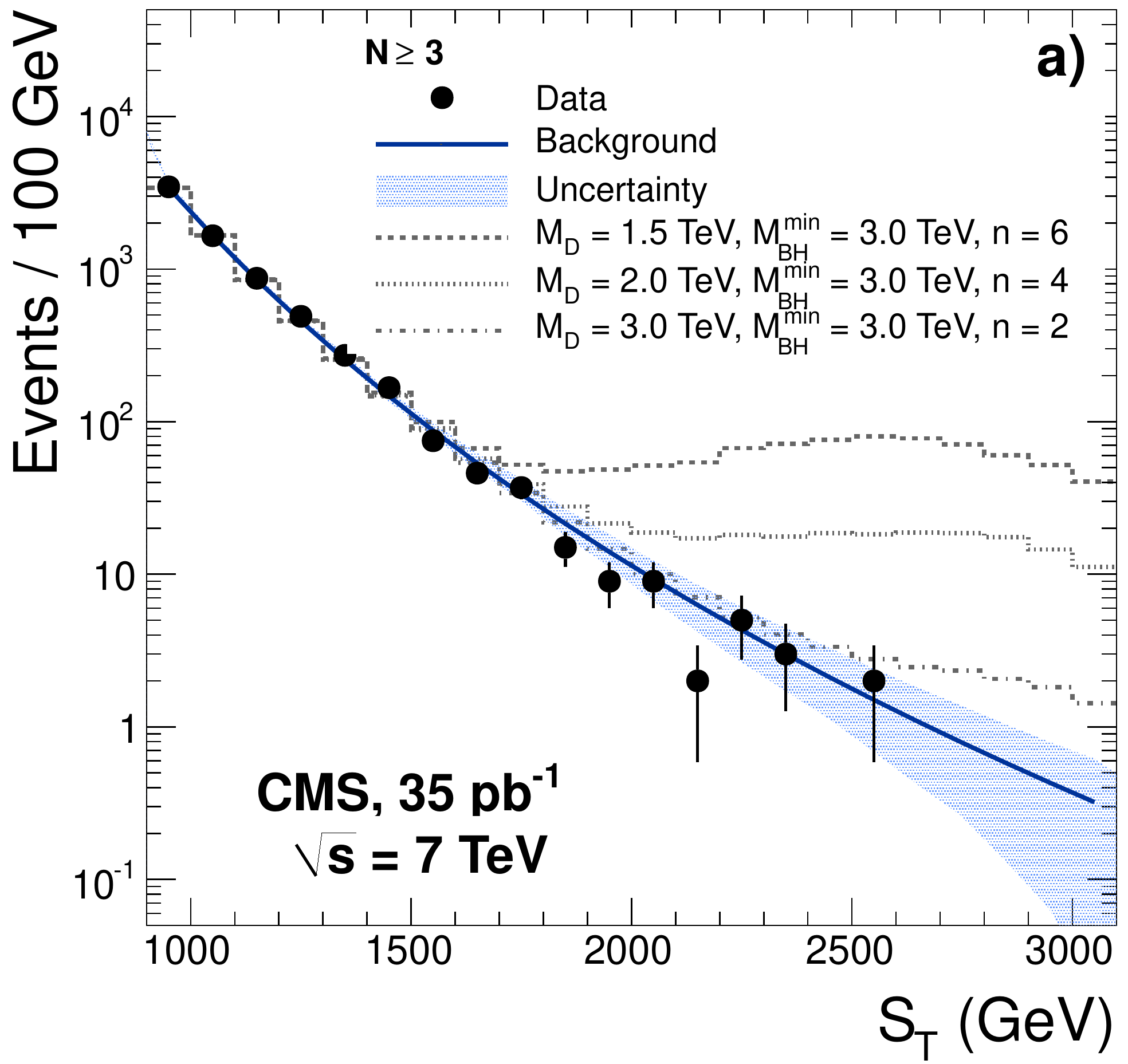}
        \includegraphics[width=.32\textwidth]{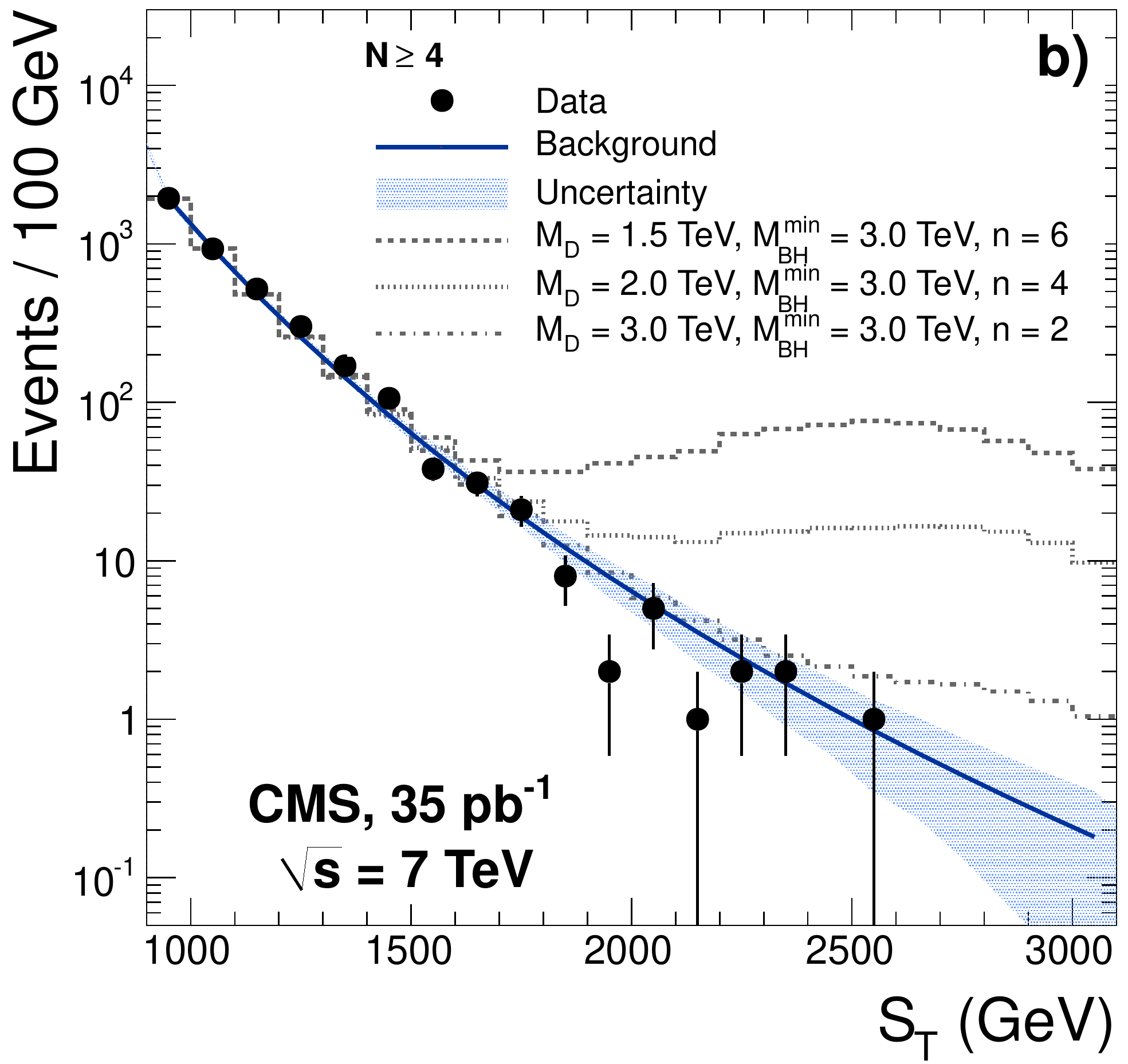}
    \includegraphics[width=.32\textwidth]{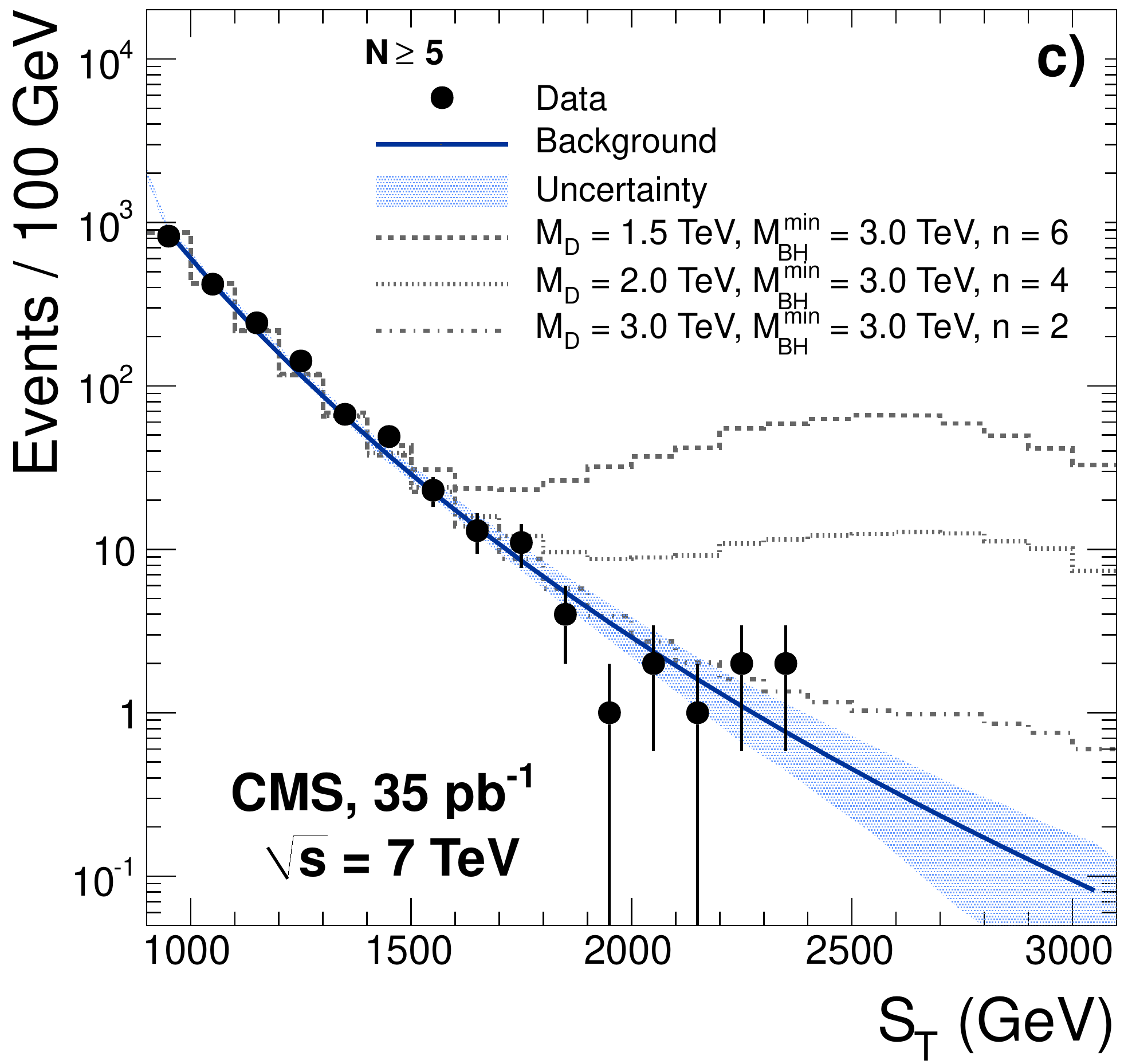}
 \end{center}
  \caption{Total transverse energy $S_T$ for events with multiplicites $n\geq3$, $n\geq4$ and $n\geq5$, respectively, obtained by the CMS using data sets with $7$ TeV and $35 {\rm fb}^{-1}$. The shaded band is the background (QCD+leptons+$t\bar{t}$) for prediction with its uncertainty. The figures are from \cite{CMS}. Similar results can be found in \cite{CMS2} for $1{\rm fb}^{-1}$ data.}
\label{fig:CMS}
\end{figure}

\begin{figure}[h]
\begin{center}
    \includegraphics[width=.33\textwidth]{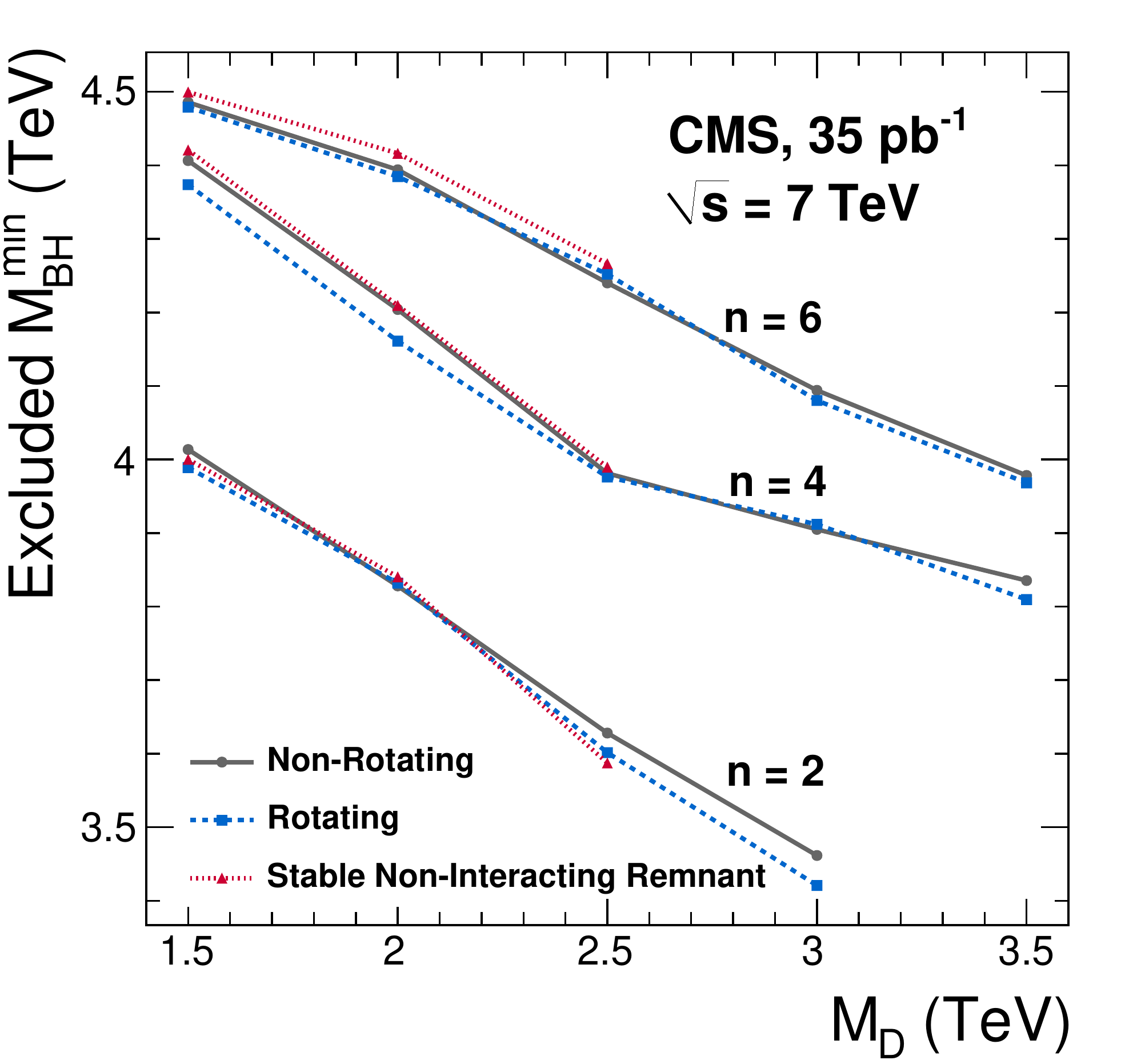}
        \includegraphics[width=.33\textwidth]{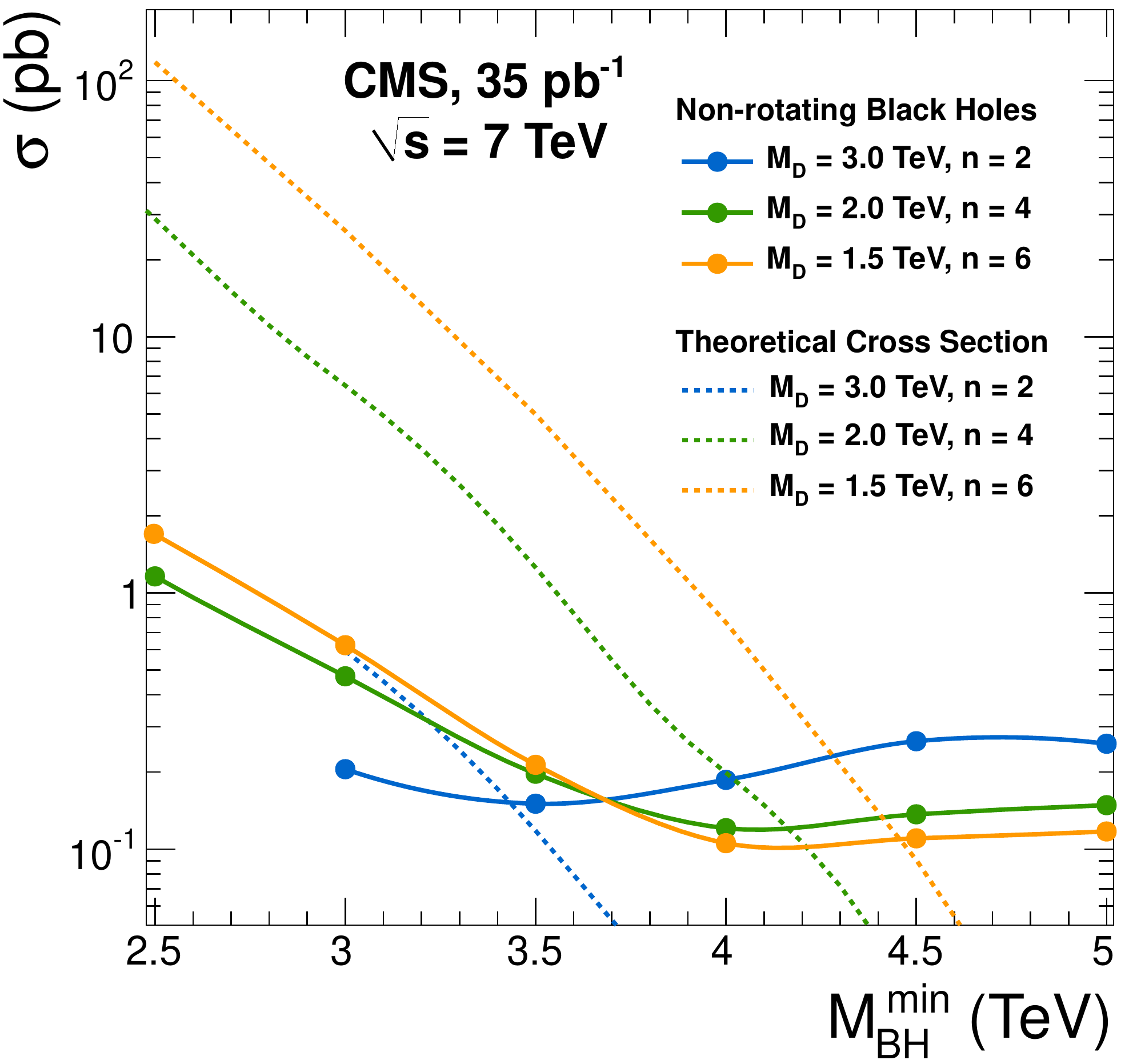}
 \end{center}
  \caption{The $95\%$ confidence level upper limits on the produced black hole cross-section and mass in  the CMS results. The figures are from \cite{CMS}. Similar results can be found in \cite{CMS2}.}
\label{fig:CMS2}
\end{figure}

The main results are depicted in Fig. \ref{fig:CMS} and Fig. \ref{fig:CMS2} (both are from \cite{CMS}). Similar but slightly improved results  with a larger data sample are found in \cite{CMS2}. In summary, no significant deviation from the background  was found in the $35~{\rm fb}^{-1}$ nor in the $1~{\rm fb}^{-1}$ data samples. Only a limit on black hole mass (see Fig. \ref{fig:CMS2}) for given values of $M_D$ in the TeV range could be set. In \cite{CMS2}, results for higher multiplicities up to $N\geq 8$ are presented, but still no significant deviation has been found.


Some comments on the recent CMS results are in order. 
As was emphasized previously \cite{CMS, CMS2}, the results should be treated as indicative, rather than precise, as the semi-classical approximation is not valid in the parameter region covered by the experiment so far \cite{Park:2011}. 
There are two main cautions in interpreting the results:
\begin{itemize}
\item The parameter range in the CMS results  is actually out of the validity range of the semi-classical approximation of a black hole.  
\item Interpretation of the results depends on particular realization of the low energy gravity models.  
\end{itemize}

 Having a limited collision energy $\sqrt{s}=7$ TeV, the considered `black hole' cannot be heavy enough beyond the quantum gravity regime in ${\cal O}(1)$ TeV. The ratio of the minimum black hole mass and the fundamental scale, which was excluded by the CMS data, is in the range \cite{CMS, Park:2011}
\begin{eqnarray}
\left(\frac{M}{M_D}\right)_{\rm CMS} &\in& \left[\frac{3.5}{3.0}, \frac{4.0}{1.5}\right]_{D=6},\\
&\in&\left[\frac{4.0}{3.5}, \frac{4.5}{1.5}\right]_{D=10},
\label{eq:CMS}
\end{eqnarray}
or the corresponding size of the Schwarzschild radius is
\begin{eqnarray}
\left(r_{BH}\right)_{\rm CMS} &\in& \left[0.9, 2.1\right]\ell_6,\\
&\in& \left[0.9,2.4\right]\ell_{10}.
\end{eqnarray}
%

\begin{figure}[h]
\centering
\includegraphics[width=0.45\textwidth]{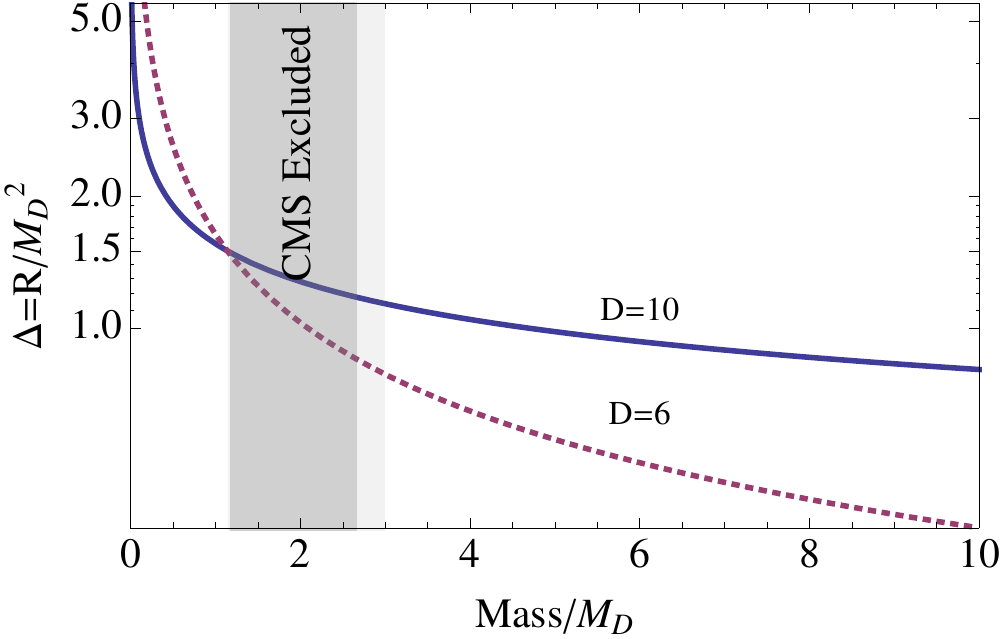}
\includegraphics[width=0.45\textwidth]{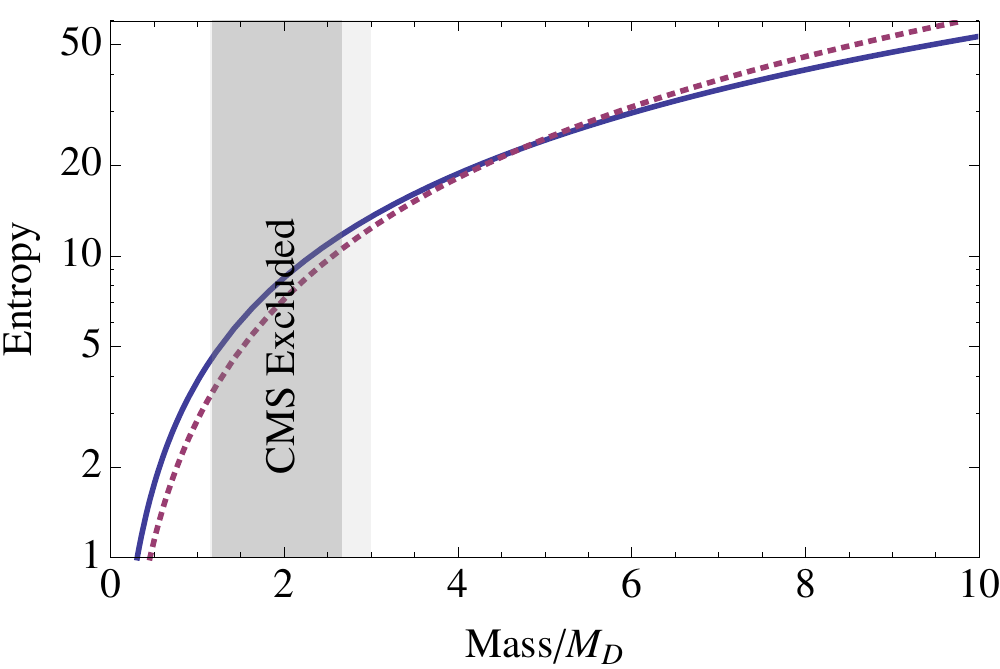}
\caption{\label{Fig:entropy} The higher order curvature term, $\Delta$, (top) and entropy (bottom) are plotted for a higher dimensional black hole ($D=6$(dotted), $D=10$(solid)).  The vertical column in gray is the CMS exclusion region: $M/M_D=[3.5/3.0 (4/3.5), 4.0/1.5(4.5/1.5)]$ for $D=6(10)$.}
\label{Fig:entropy}
\end{figure}

As was discussed in earlier sections, the validity of semi-classical approximation is guaranteed when the black  hole is massive enough, $M\gg \md$. However, in the claimed exclusion range, the large quantum gravitational correction of the order of $\Delta = \left(\frac{\ell_D}{r_s(\sqrt{s})}\right)^p \gsim 100\%$ with a positive (unknown) power $p>0$ is expected. We may not impose any geometrical meaning on an object with a small entropy.  See an estimation of higher order correction (upper) and the extrapolated value of entropy (lower) in Fig. \ref{Fig:entropy} for $D=6$ (dotted lines) and $D=10$ (solid lines). For consistency of the perturbative expansion, the correction term must be significantly smaller than the leading order term ($\Delta \ll 1$) and entropy should be large ($S \gg 1$), but in the CMS exclusion range (vertical columns in gray scales), the correction is expected to be as large as the leading term. The entropy of the black hole is ${\cal O}(1)$.  Within this parameter space, all the MC simulations suffer from large quantum corrections, which can lead a significant change in the final result. There may be a chance for the LHC  to form some black hole precursors  \cite{Dimopoulos:2001st, Meade}  and low energy Kaluza-Klein excitation modes of bulk fields.\footnote{An important counter example is the model in the presence of a compact hyperbolic extra dimension background where  no low energy excited Kaluza-Klein modes are expected (see Sec. \ref{Sec:CHS}).}  However, a semi-classical black hole now seems out of the reach of the LHC run with $\sqrt{s}=7$ TeV.


\subsection{A guide for future improvement }

Here is a guide for future improvement \cite{Cardoso:2005aaa}.
\begin{itemize}
  \item The detailed brane realization of the standard model is still unknown (see Sec. \ref{sec:HighD}).  

  \item The balding phase should be understood through dynamical simulation, which can  most probably only be done via a numerical study (see Sec. \ref{sec:numerical}).
  \item For $D\gg4$, spin-2 graviton emission to the bulk from highly rotating black holes can be sizable and dominant.
  Hawking radiation to a spin-2 particle has been considered only for the non-rotating case in higher dimensions $D>4$ ~\cite{graviton1,graviton2, graviton3}.
  \item The final state of a black hole, i.e., the Planck phase,  is extremely poorly understood. Full quantum gravitational consideration is required (see e.g. ~\cite{maldacena,final1,final2,final3,final4} for some recent discussions).
  \item Details of signals should depend on many factors (e.g., 2 to 2 dominance ~\cite{Meade}, inelastic effect~\cite{inelastic}, recoil~\cite{Frolov:2002, Frolov:2002b, Frolov:2004a, Frolov:2004b}, split-brane~\cite{rsadd, split}, hidden symmetry \cite{Frolov:2006hd, Frolov:2008hd},  instability in relation with quasi-normal mode and black hole bomb \cite{Berti:2009kk, Berti:2004ju} and \cite{BB, BB2,Rosa:2009},  etc.).
\item The quantum gravity correction to black hole geometry and Hawking radiation should be modeled; see \cite{Konoplya:2010} (Gauss-Bonnet corrected Black hole) and \cite{Nicolini:2011} (non-commutative geometry inspired black hole).

\end{itemize}


\section{Conclusions  \label{sec:conclusion}}

The LHC is the most promising place to look for new physics models on the TeV scale including the low scale gravity models such as ADD and RS. If one of the low scale gravity models is realized in nature, the microscopic black holes on the 1 TeV scale will be produced at the LHC and their signatures, especially the Hawking radiation, might provide a valuable guide to providing a deeper level of understanding of quantum gravity and spacetime structure. 
In this article, we review the progress of understanding this exciting area of research over the past decade, emphasizing both theoretical and experimental efforts. 
On the theoretical side, analytic as well as numerical studies have been performed to understand the dynamical process of black hole formation in high energy collisions mostly in the semi-classical domain. The Hawking radiation to brane localized fields (and also partly to the bulk fields) from a higher dimensional rotating black hole is  understood via the precise calculation of greybody factors. Monte-Carlo simulation codes have been developed for the ``black hole events'' at the LHC.  On the experimental side, the LHC begins to produce data for the TeV scale gravity models and will reach the true trans-Planckian or microscopic black hole domain in the coming years after its collision energy is upgraded to $14$ TeV from the current $7$ TeV as the LHC was originally designed. Even though the current understanding is limited due to the lack of a full quantum gravity model, we hope to learn more by observing real data from the TeV domain.

Finally, a short comment on the safety of the LHC is in order. The recent analysis shows that the macroscopic effects of TeV-scale black holes should have already been seen in various astrophysical environments if those black holes are dangerously long-lived or even stable \cite{safety1,safety2}. This analysis is important since it excludes the possibility that any mini black holes that might be produced by the LHC could cause catastrophic damage.  If the LHC can produce black holes, there are various places in the universe such as neutron stars and white dwarfs where similar mini black holes can also be produced  at an even more frequent rate by collisions of the ultra-high energy cosmic ray particles and nucleons in those dense astronomical objects.
Basically all  of those dense astrophysical objects are of the same or a higher level of danger. Thus, the null observed signals from those dense astrophysical objects ensures that there is no risk of any significance whatsoever from such black holes.

\newpage
\section*{Acknowledgements}

This research was supported by the Basic Science Research Program through the National Research Foundation of Korea (NRF) funded by the Ministry of Education, Science and Technology (2011-0010294) and (2011-0029758) and also by Chonnam National University. The early phase of the work was done at Institute for the Physics and Mathematics of the Universe (IPMU), Tokyo University. At `HEP/NR workshop' in Madeira on 31 Aug. -3  Sep. 2011 (see a summary of the workshop in Ref.~\cite{Madeira}), I had very helpful discussions with  participants, especially with Steve Giddings, Andy Parker, Greg Landsberg,  Masaru Shibata, Roberto Emparan, Frans Pretorius, Vicki Moeller, Hirotada Okawa, Akihiro Ishibashi , V\'itor Cardoso,  Harvey Reall, Ruth Gregory, Veronika Hubeny, Valeria Ferrari, Toby Wiseman, \'Oscar Dias, Carlos Herdeiro, Leonardo Gualtieri, Marco Sampaio, Miguel Zilh\~ao, Luis Lehner, Nicolas Yunes and Ulich Sperhake.  Finally, I deeply appreciate valuable collaborations and discussions with Kin-ya Oda, Daisuke Ida, Hirotaka Yoshino, James Frost, Domenico Orlando, Yoonbai Kim, Shinji Mukohyama, Shigeki Sugimoto and Brian Webber.

\newpage

\appendix 

\numberwithin{equation}{section}

\section{Appendix: Planck units in $D$ dimensions \label{Appendix:1}}

The Planck units in $D=4+n$ dimensions can be determined as:
\begin{eqnarray}
\left[G_D\right]&=&L^{D-1}T^{-2}M^{-1},\\
\left[\hbar\right]&=&L^2 T^{-1}M,\\
\left[c\right]&=&LT^{-1},\\
\left[G_D^\alpha \hbar^\beta c^\gamma\right]&=&L^{\alpha(D-1)+2\beta+\gamma}T^{-2\alpha -\beta-\gamma}M^{-\alpha +\beta -\gamma},\\
\ell_D&\sim& \left(\frac{G_D\hbar}{c^3}\right)^{\tfrac{1}{D-2}},\\
t_D&\sim&\left(\frac{G_D\hbar}{c^{D+1}}\right)^{\tfrac{1}{D-2}},\\
\md&\sim&\left(\frac{\hbar^{D-3}c^{5-D}}{G_D}\right)^{\tfrac{1}{D-2}}.
\end{eqnarray}

The Schwarzschild-Tangherlini black radius ($r_s$) corresponding to the mass $M$ and impact parameter of the scattering $b$ in terms of angular momentum:
\begin{eqnarray}
&&r_s=C_D \left(\frac{G_D M}{c^4}\right)^{\tfrac{1}{D-3}},\\
&&b=2 \frac{J}{\sqrt{s}},
\end{eqnarray}
where
$C_D=\left(\frac{16\pi}{(D-2)\Omega_{D-2}}\right)^{\tfrac{1}{D-3}}$ and $\Omega_{D-2}=\frac{2\pi^{(D-3)/2}}{\Gamma((D-3)/2)}$.

At the limit $M=\sqrt{s}\gg M_D$ with $\hbar \to 0$ and fixed $G_D$, 
\begin{eqnarray}
&&\ell_D \to 0, \\
&&\lambda_{\rm dB} = \frac{4\pi \hbar}{\sqrt{s}} \to 0, \\
&&r_s \to {\rm finite},
\end{eqnarray}
such that one can conclude that in the trans-Planckian domain, the dynamics would be described by $r_s \gg \ell_D, \lambda_{\rm dB}$.

\section{Appendix: Solving the generalized Teukolsky equation \label{Appendix:2}}

\begin{figure}[h]
\centering
\includegraphics[width=0.8\textwidth]{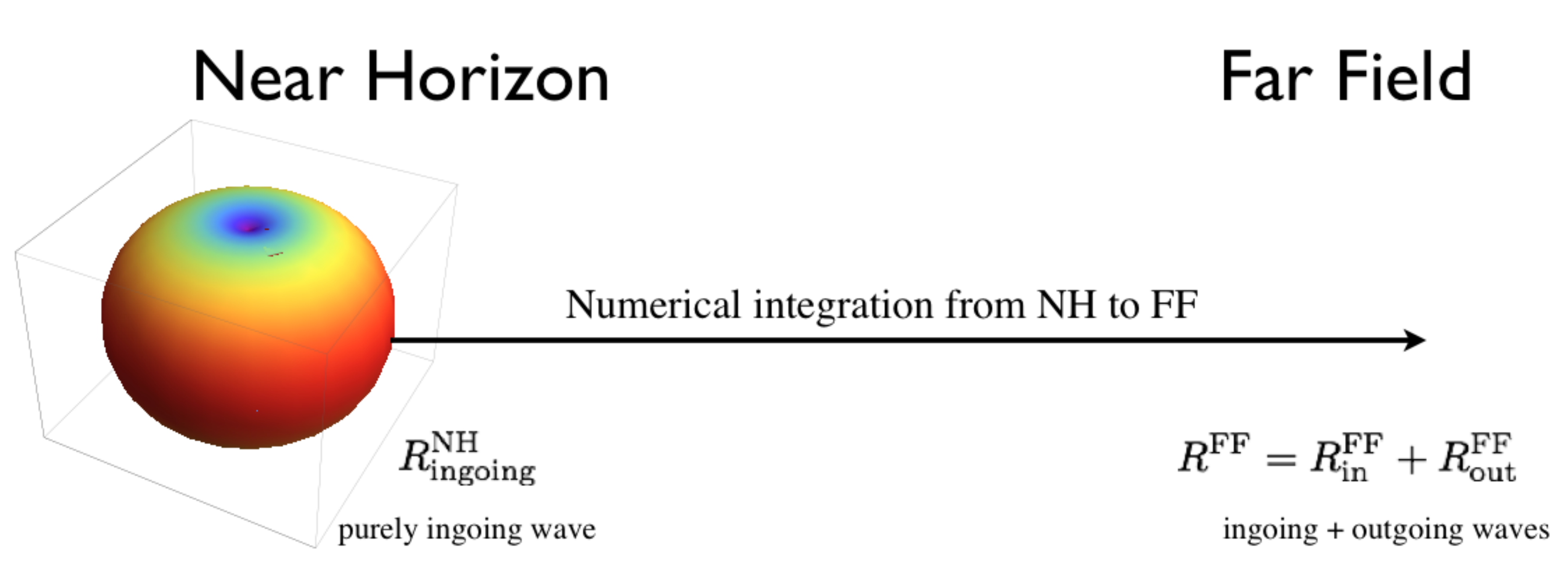}
\caption{\label{fig:numeric} Schematic summary of the method to find the greybody factor 
by solving the generalized Teukolsky equation (details in \cite{IOP3}).}
\end{figure}

The solution to the generalized Teukolsky equation can be found numerically \cite{IOP1,IOP2-1, IOP2-2,IOP3}. To get the solution, one should impose the `purely ingoing wave' condition near the horizon ($r\to r_h$) then numerically integrate the differential equation to the far field regime ($r\to \infty$) where ingoing and outgoing waves are superposed (see Fig. \ref{fig:numeric}). 

On the $(3+1)$ dimensional `brane', the induced metric of the $(4+n)$-dimensional
Myers-Perry solution with a single nonzero angular momentum (on the brane)  is given by
\begin{eqnarray}
-ds^2&=&{\Delta-a^2\sin^2\vartheta\over\Sigma}dt^2
+{2a(r^2+a^2-\Delta)\sin^2\vartheta\over\Sigma}dtd\varphi\nonumber\\
&&{}-{(r^2+a^2)^2-\Delta a^2\sin^2\vartheta\over\Sigma}\sin^2\vartheta d\varphi^2
-{\Sigma\over\Delta}dr^2-\Sigma d\vartheta^2,
\label{myers-perry}
\end{eqnarray}
where
\begin{eqnarray}
\Sigma=r^2+a^2\cos^2\vartheta,~~~
\Delta=r^2+a^2-\mu r^{1-n}.
\end{eqnarray}
The parameters $\mu$ and $a$ are equivalent to the total mass $M$ and
the angular momentum $J$ evaluated at the spatial infinity of the $(4+n)$-dimensional space-time (see Sec. \ref{sec:blackobjects}). 

The radial equation of the generalized Teukolsky equation is given by a linear, second-order equation as shown in \cite{IOP1}:
\begin{eqnarray}
\frac{d^2 R}{d r^2}+{\cal \eta} \frac{d R}{d r} +{\cal \tau} R =0,
\label{eq:diffeqKN}
\end{eqnarray}
where $\eta$ and $\tau$ are determined in the Kerr-Newman frame as:
\begin{eqnarray}
\eta=-\frac{(s-1)\Delta'+2iK}{\Delta}, 
\tau=\frac{2i \omega r(2s-1)-\lambda}{\Delta}
\end{eqnarray}
where $K=(r^2+a^2)w - m a$.

\subsection{Near horizon: $r\to r_h$} 

To find the near horizon (NH) solution, it is convenient to expand all quantities in the $(r-r_h)^k$ series:
\begin{eqnarray}
\Delta =\sum_{k=1} \Delta_k (r-r_h)^k, \\
K=\sum_{k=1} K_k (r-r_h)^k,\\
\eta = \sum_{k=-1} \eta_k (r-r_h)^k,\\
\tau =\sum_{k=-1} \tau_k (r-r_h)^k.
\end{eqnarray}

Then the solutions near the horizon (NH)  $r\rightarrow r_h$ are read
\begin{eqnarray}
R^{\rm NH}_{\rm in} &=& 1+ a_1 (r-r_h) + \frac{a_2}{2}(r-r_h)^2+ \cdots, \nonumber \\
R^{\rm NH}_{\rm out} &=& e^{2 i k r_*}(r-r_h)^s\left(1+ b_1(r-r_h)
+\cdots \right),
\end{eqnarray}
where the expansion coefficients $a_i$'s and $b_i$'s are found as follows:
\begin{eqnarray}
&&a_1=-\frac{\tau_{-1}}{\eta_{-1}}, \\
&&a_2= -\frac{(\eta_0+\tau_{-1})a_1+\tau_0}{1+\eta_{-1}},\\
&&\cdots.
\end{eqnarray}

The problem now is to integrate Eq.\eqref{eq:diffeqKN} from purely ingoing
initial conditions at $r=r_h$ out to $r\rightarrow \infty$.  However, $R^{\rm NH}_{\rm out}$ 
is unstable against contamination of the purely ingoing solution. To counteract the above contamination, let
\begin{eqnarray}
\tilde{R}= R-\left(1+ a_1 (r-r_H)\right) \approx \frac{a_2}{2} (r-r_H)^2+\cdots
\end{eqnarray}
in the vicinity of the horizon, which satisfies the following equation:
\begin{eqnarray}
{\cal L} \tilde{R} = g, \label{eq:diffeq2}
\end{eqnarray}
where ${\cal L}=d^2/dr^2 +\eta d/dr +\tau$ and $g=-{\cal L}\left(1+
a_1 (r-r_H)\right)=-\eta a_1-\tau\left(1+ a_1 (r-r_H)\right)$.
Equation \eqref{eq:diffeq2} can be integrated  numerically from the near horizon to the far field
regime. Calculation details could be found in \cite{IOP3}.

\subsection{Far field limit : $r\to \infty$} 
At the far field limit (FF), the solutions to the radial equation can be found easily.

For a spinor field with spin $s=1/2$,

\begin{eqnarray}
R^{\rm FF}_{\rm in} &\sim& 1+ \frac{c^f_1}{r} + \frac{c^f_2}{r^2}+\frac{c^f_3}{r^3} +\cdots , \nonumber \\
R^{\rm FF}_{\rm out} &\sim& e^{2 i k r_*}\frac{1}{r} ( 1+
\frac{d^f_1}{r}+\frac{d^f_2}{r^2}\cdots),
\end{eqnarray}
where
\begin{eqnarray}
c^f_1&=&-i \frac{\lambda}{2\omega},\\
\cdots
\end{eqnarray}
Thus  $\tilde{R}_{1/2}$ is read
\begin{eqnarray}
\tilde{R}_{1/2}(r\rightarrow \infty) \simeq \left(Y_{\rm in}+1-a_1
\right) + a_1 r+\frac{Y_{\rm in} c^f_1}{r}+Y_{\rm
out}\frac{e^{2i\omega r_*}}{r}.
\end{eqnarray}

Using this expression, we can easily read out $Y$s without
numerical difficulties at FF.
The greybody factor is found as follows:
\begin{eqnarray}
\Gamma_{s=1/2} = 1-\frac{2\omega}{|c^f_1|}\frac{|Y_{\rm
out}|^2}{|Y_{\rm in}|^2}.
\end{eqnarray}

For a vector field with spin $s=1$,
\begin{eqnarray}
R^{\rm FF}_{\rm in} &\sim& r (1+ \frac{c^v_1}{r} + \frac{c^v_2}{r^2}+\frac{c^v_3}{r^3} +\cdots) , \nonumber \\
R^{\rm FF}_{\rm out} &\sim& e^{2 i k r_*}\frac{1}{r} ( 1+
\frac{d^v_1}{r}+\frac{d^v_2}{r^2}+\frac{d^v_3}{r^3} +\cdots),
\end{eqnarray}
where
\begin{eqnarray}
c^v_1&=&-i \frac{\lambda}{2\omega},\\
c^v_2&=&-\frac{\lambda^2-4 a\omega (a\omega-m)}{8\omega^2},\\
&\cdots&.\nonumber
\end{eqnarray}
Then, $\tilde{R}_1$ is read
\begin{eqnarray}
\tilde{R}_1(r\rightarrow \infty) \simeq \left(Y_{\rm in}-a_1
\right)r + \left(Y_{\rm in}c^v_1-1+a_1\right)+\frac{Y_{\rm in}
c^v_2}{r}+Y_{\rm out}\frac{e^{2i\omega r_*}}{r}.
\end{eqnarray}

Similarly when $s=1/2$, one can easily read out $Y$s without
numerical difficulties at FF. Finally, the greybody factor for a vector field can be determined as
\begin{eqnarray}
\Gamma_{s=1} = 1-\frac{2\omega^2}{|c^v_2|}\frac{|Y_{\rm
out}|^2}{|Y_{\rm in}|^2}.
\end{eqnarray}

\end{document}